\documentclass[twocolumn]{aastex631}
\usepackage{amsmath}     
\usepackage{wasysym}           
\usepackage{graphicx}
\usepackage{amssymb}
\usepackage{epstopdf}
\usepackage{mathrsfs}
\usepackage{hyperref}
\usepackage{anyfontsize}
\usepackage{natbib}
\usepackage{color}
\usepackage{lipsum}
\usepackage{diagbox}

\shorttitle{Repeating nuclear transients}
\shortauthors{Bandopadhyay et al.}
\begin{document}
\title{Repeating nuclear transients from repeating partial tidal disruption events: reproducing ASASSN-14ko and AT2020vdq}
\author[0000-0002-5116-844X]{Ananya Bandopadhyay}
\affiliation{Department of Physics, Syracuse University, Syracuse, NY 13210, USA}

\author[0000-0003-3765-6401]{Eric R.~Coughlin}
\affiliation{Department of Physics, Syracuse University, Syracuse, NY 13210, USA}

\author[0000-0002-2137-4146]{C.~J.~Nixon}
\affiliation{School of Physics and Astronomy, Sir William Henry Bragg Building, Woodhouse Ln., University of Leeds, Leeds LS2 9JT, UK}

\author[0000-0003-1386-7861]{Dheeraj R.~Pasham}
\affiliation{MIT Kavli Institute for Astrophysics and Space Research, Cambridge, MA 02139, USA}

\email{abandopa@syr.edu}
\email{ecoughli@syr.edu}

\begin{abstract}
Some electromagnetic outbursts from the nuclei of distant galaxies have been found to repeat on months-to-years timescales, and each of these sources can putatively arise from the accretion flares generated through the repeated tidal stripping of a star on a bound orbit about a supermassive black hole (SMBH), i.e., a repeating partial tidal disruption event (rpTDE). Here we test the rpTDE model through analytical estimates and hydrodynamical simulations of the interaction between a range of stars, which differ from one another in mass and age, and an SMBH. We show that higher-mass ($\gtrsim 1 M_{\odot}$), evolved stars can survive many ($\gtrsim 10-100$) encounters with an SMBH while simultaneously losing $few \times 0.01 M_{\odot}$, resulting in accretion flares that are approximately evenly spaced in time with nearly the same amplitude, quantitatively reproducing ASASSN-14ko. We also show that the energy imparted to the star via tides can lead to a change in its orbital period that is comparable to the observed decay in the recurrence time of ASASSN-14ko's flares, $\dot{P}\simeq-0.0026$. Contrarily, lower-mass and less-evolved stars lose progressively more mass and produce brighter accretion flares on subsequent encounters for the same pericenter distances, leading to the rapid destruction of the star and cessation of flares. Such systems cannot reproduce ASASSN-14ko-like transients, but are promising candidates for recreating events such as AT2020vdq, which displayed a second and much brighter outburst compared to the first. Our results imply that the lightcurves of repeating transients are tightly coupled with stellar type. 
\end{abstract}


\section{Introduction}
\label{sec:intro}
Observations suggest that the centers of almost all galaxies contain SMBHs \citep{kormendy13}. When a star in a galactic nucleus is placed onto a low angular momentum orbit about an SMBH and passes within a critical distance of it, the tidal forces of the SMBH can either partially or completely disrupt the star. Such a destructive encounter between a star and an SMBH is known as a tidal disruption event (TDE) (e.g., \citealt{rees88, gezari21}). The accretion of the tidally disrupted debris from the star onto the SMBH generates a luminous flare. With the increased cadence and depth of time-domain surveys, such as the All Sky Automated Survey for Supernovae (ASAS-SN)~\citep{shappee14}, the Zwicky transient facility (ZTF)~\citep{bellm14}, and the Dark Energy Survey \citep{DESI16}, tens of TDEs are being observed every year (e.g., \citealt{arcavi14, holoien14,vanvelzen16, gezari17, pasham18, vanvelzen21, payne21, wevers21, lin22, nicholl22, guolo23, hammerstein23, pasham23, wevers23, yao23}), a number that is anticipated to increase by at least an order of magnitude in the coming 1-2 years, when the Vera Rubin Observatory becomes operational~\citep{ivezic19,bricman20}. 

The standard TDE lightcurve is predicted to rise, peak, and decay monotonically with time $t$, and while the initial rise and peak are dependent on a number of factors involving the structure of the star (e.g., \citealt{lodato09, guillochon13, golightly19a, lawsmith20, jankovic23}), the asymptotic decline of the accretion rate with time scales either as $\propto t^{-5/3}$ if the disruption is complete \citep{rees88, phinney89} or as $\propto t^{-9/4}$ if the disruption is partial \citep{coughlin19}. Recently, however, there have been TDE flares\footnote{Quasi-periodic eruptions (QPEs; e.g., \citealt{miniutti19, giustini20, arcodia21, chakraborty21, quintin23, arcodia24, pasham24}) are qualitatively similar in terms of their highly cyclic behavior but are exclusively seen in the X-ray and have significantly shorter timescales (but see \citealt{evans23, guolo24}), and while there may be similarities in their physical origin that are related to TDEs, we restrict our phenomenological focus to longer-duration  repeating events.} -- including ASASSN-14ko \citep{payne21}, AT2018fyk \citep{wevers23}, eRASSt-J045650 \citep{liu24}, and AT2020vdq \citep{somalwar23} -- that rebrighten on timescales of months to years following the initial peak, thus challenging this classic picture. One  interpretation is that these sources represent repeating partial TDEs (rpTDEs), in which a star is orbiting an SMBH on a highly eccentric orbit, the pericenter distance of which is comparable to the partial disruption radius of the star; the latter is $\sim 2 r_{\rm t}$, where $r_{\rm t} = R_{\star}\left(M_{\bullet}/M_{\star}\right)^{1/3}$ is the standard tidal disruption radius \citep{hills75} with $R_{\star}$ the stellar radius and $M_{\bullet}$ and $M_{\star}$ the SMBH and stellar mass, respectively.

Among the few rpTDE candidates, the longest-known and most heavily studied candidate is ASASSN-14ko, which has flared $\gtrsim 20$ times since its initial detection \citep{holoien17,kochanek17,payne21}. The energetics of the outbursts are consistent with the stripping of a star in which $\sim few\times 0.01M_{\odot}$ of mass is liberated -- and subsequently accreted -- during each pericenter passage if the luminosity is $L = 0.1\dot{M}c^2$  \citep{cufari22b}. If ASASSN-14ko formed approximately contemporaneously with the first detection, meaning that there were no outbursts prior to the original detection (which must be the case if the star is near-solar, as it will only survive $\lesssim 100$ encounters if the per-encounter stripped mass is $\sim 0.01 M_{\odot}$), then the orbital period of the star of $\sim 115$ days can be explained if the star was tidally captured from a tight binary \citep{cufari22b}. However, if the star loses $\sim 0.001 M_\odot$ on each encounter, with the mass stripped per-encounter increasing gradually with time, then the system could have existed for many more orbital periods prior to the original detection, and there are a number of mechanisms, including gravitational-wave inspiral and the interaction with a disc~\citep{linial24}, that can then explain its current orbital period, though another mechanism must then be responsible for perturbing its pericenter distance to within $\sim 2 r_{\rm t}$.

While the rpTDE model has been adopted to explain periodic nuclear transients, it has yet to be rigorously tested with hydrodynamical simulations. \cite{antonini11} performed smoothed particle hydrodynamics simulations to study the repeated partial disruption of stars as one of the possible outcomes of three-body interactions between SMBHs and stellar binaries. However, the detailed hydrodynamical modeling of lightcurves from these sources, which includes resolving the rate of return of tidally disrupted debris to the SMBH after many encounters, has not been explored. One of the potential issues of this model is the spin imparted to the star: the star is torqued in a prograde sense following its interaction with the SMBH, which renders it more susceptible to disruption \citep{golightly19a}. If net angular momentum is imparted to the star with each pericenter passage, it may only be able to survive a few encounters before being completely destroyed, thus precluding the model from describing events -- such as ASASSN-14ko -- that have flared many times. 

Here we study the hydrodynamical evolution of a star that is being repeatedly stripped of mass by an SMBH, both to understand the stellar survivability and the viability of the rpTDE model. In Section~\ref{sec:lagrangian} we construct a simple lagrangian toy model to estimate the angular momentum imparted to the star through tidal interactions with the SMBH, and we derive the induced change in the orbital period and pericenter distance of the star. In Section~\ref{sec:hydro} we present the results of hydrodynamical simulations of the repeated partial disruption of stars by an SMBH. We show that an ideal candidate for producing repeated flares, the strength and recurrence of which are comparable to the ASASSN-14ko event, is a massive ($\gtrsim 1M_{\odot}$) and evolved star, on an orbit having its pericenter distance roughly equal to the tidal radius of the star, $r_{\rm p} \sim r_{\rm t}$. We show that this type of star can lose $\sim 0.01 M_{\odot}$ at each pericenter passage, and survive multiple encounters, giving rise to ASASSN-14ko-like flares. We also show that by varying the stellar type and the impact parameter characterizing the distance of closest approach between the star and the SMBH, the rpTDE model can be used to qualitatively reproduce the lightcurves of other transients, including AT2018fyk and AT2020vdq. We discuss our main results and implications in the context of observations in Section~\ref{sec:conclusions} before summarizing in Section~\ref{sec:summary}.

\section{Toy Problem}
\label{sec:lagrangian}
One of the main potential issues with the rpTDE model is the imparted rotation to the stellar core: upon initially and strongly interacting tidally with the SMBH, the star will be spun up to a rotational velocity that can -- depending on the depth of the encounter and the properties of the star -- be a substantial fraction of the stellar break-up velocity, $\Omega_{\star} \simeq \sqrt{GM_{\star}}/R_{\star}^{3/2}$. If the star is spun up by a similar amount on each encounter, it will be rapidly (i.e., after only a few encounters) destroyed, precluding the possibility of producing, e.g., ASASSN-14ko, which has now flared $N\approx 20$ times. 

However, the notion that the star is spun up repeatedly is generally not correct\footnote{\label{footnote:note1}A secular increase in the rotational velocity could be achieved if the frequency of one of the stellar eigenmodes occurs in resonance with the orbital frequency, but it is difficult to see how changes in the stellar structure (induced by the rotation itself) would not cause the star to move off this resonance after only a few repeated encounters.}, because the star needs to be rotating at a rate that is smaller than the rotational velocity at pericenter in order to be efficiently torqued by the SMBH. Once the star achieves an angular velocity that is comparable to $\Omega_{\rm p} \simeq \sqrt{GM_{\bullet}\left(1+e\right)}/r_{\rm p}^{3/2}$, where $r_{\rm p} \simeq few\times r_{\rm t}$ is the pericenter distance, the SMBH appears stationary (near pericenter) in the corotating frame of the star. In this case, then, there is no relative motion between the stellar surface and the speed with which the SMBH moves in the local frame of the star, thus preventing successive and efficient gravitational torques on the stellar body.

Once the star is spun up to $\sim \Omega_{\rm p}$ and if there is no viscous decay of the oscillatory quadrupole moment of the star, we expect there to be a pseudo-random change in the angular velocity of the star during successive encounters, owing simply to the fact that there is a time-dependent (on the timescale of the dynamical time of the star, modulo the eigenvalue appropriate to roughly the $f$-mode of the progenitor, which predominantly contributes to the tidal excitation of the star; e.g., \citealt{press77}) phase of the $\ell = 2$ oscillation of the star. Thus, if the oscillatory modes of the progenitor do not damp radiatively (or viscously, in the presence of another form of viscosity that could arise from, e.g., magnetic effects) by the time of the next pericenter passage, there will be an inherent and stochastic coupling between the tidal field of the black hole and the $\ell = 2$ mode of the star, resulting in small changes to the angular velocity about the mean value of $\Omega_{\rm p}$ (modulo the point raised in footnote \ref{footnote:note1}). 

To illustrate this effect -- that the star can only be efficiently spun up to $\Omega_{\rm p}$ -- consider the very simple toy model of two objects, each of mass $M_{\star}/2$, connected by a rigid rod of length $2R_{\star}$ that approach a black hole of mass $M_{\bullet} \gg M_{\star}$. Denote the distance of the center of mass (COM) of this ``dumbbell'' by $r_{\rm c}$, the angle that the COM makes with the pericenter (of the COM) to the SMBH by $\phi_{\rm c}$, and the angle that one of the objects makes about the COM by $\varphi$, and further assume that the plane of rotation of the dumbbell is coincident with the plane occupied by the COM. Then the Lagrangian describing the dynamics of the system is
\begin{eqnarray}
    \mathscr{L} &=& \dot{r}_{\rm c}^2+r_{\rm c}^2\dot{\phi}_{\rm c}^2+\dot{\varphi}^2+\left(1+r_{\rm c}^2-2r_{\rm c}\cos\left[\phi_{\rm c}-\varphi\right]\right)^{-1/2}  \notag \\ 
    &+&\left(1+r_{\rm c}^2+2r_{\rm c}\cos\left[\phi_{\rm c}-\varphi\right]\right)^{-1/2}. \label{lagfull}
\end{eqnarray}
Here dots denote differentiation with respect to $\tau$, where $\tau = t\sqrt{GM_{\bullet}}/R_{\star}^{3/2}$, and $r_{\rm c}$ is measured in units of $R_{\star}$, i.e., we have let $r_{\rm c} \rightarrow r_{\rm c}/R_{\star}$. From this the Euler-Lagrange equations can be constructed that describe the motion of the dumbbell; we can also make the tidal approximation and expand the radicals in Equation \eqref{lagfull} to the quadrupole order in $1/r_{\rm c}$. 
However, the full form given in Equation \eqref{lagfull} must strictly conserve the total energy of the system, and also accounts for the back-reaction of the imparted rotational energy to the dumbbell on the energy of the COM. Therefore, it is simpler (and more accurate) to deal with the exact Lagrangian (Equation \ref{lagfull}) rather than the tidally approximated one.

\begin{figure*}
    \includegraphics[width=0.995\textwidth]{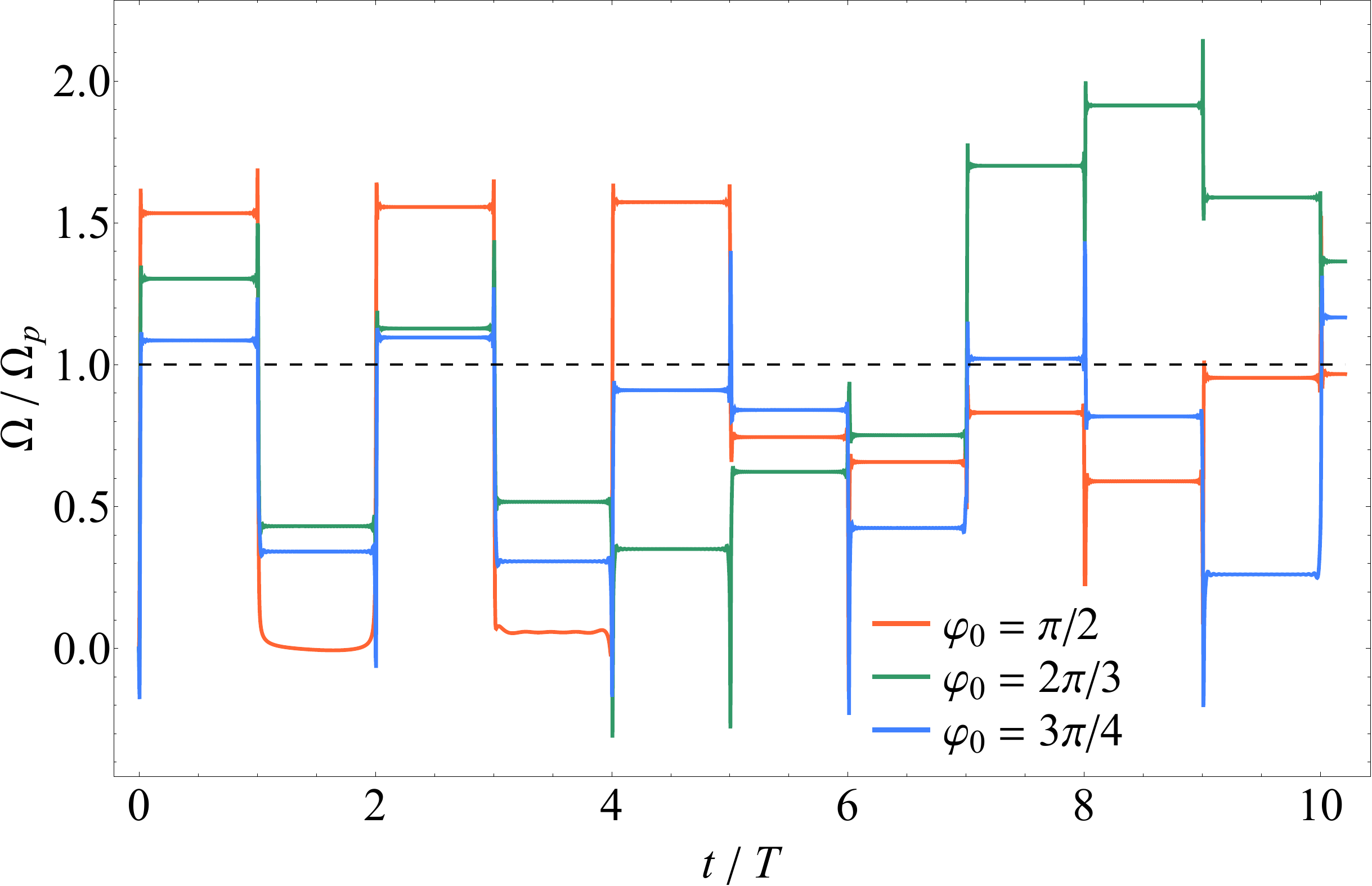}
    \caption{The angular velocity of the dumbbell relative to $\Omega_{\rm p} = \sqrt{GM_{\bullet}\left(1+e\right)}/r_{\rm p}^{3/2}$ as a function of time in units of the Keplerian orbital period, $T$. The different curves correspond to the initial phase of the dumbbell with respect to the argument of pericenter. This figure demonstrates that the dumbbell is originally spun up to an angular velocity that is comparable to $\Omega_{\rm p}$, and on each successive pericenter passage the dumbbell receives a pseudo-random kick to its angular velocity, resulting in a scatter about the mean value of 1.}
    \label{fig:Omega_of_t}
\end{figure*}

\begin{figure*}
    \includegraphics[width=0.485\textwidth]{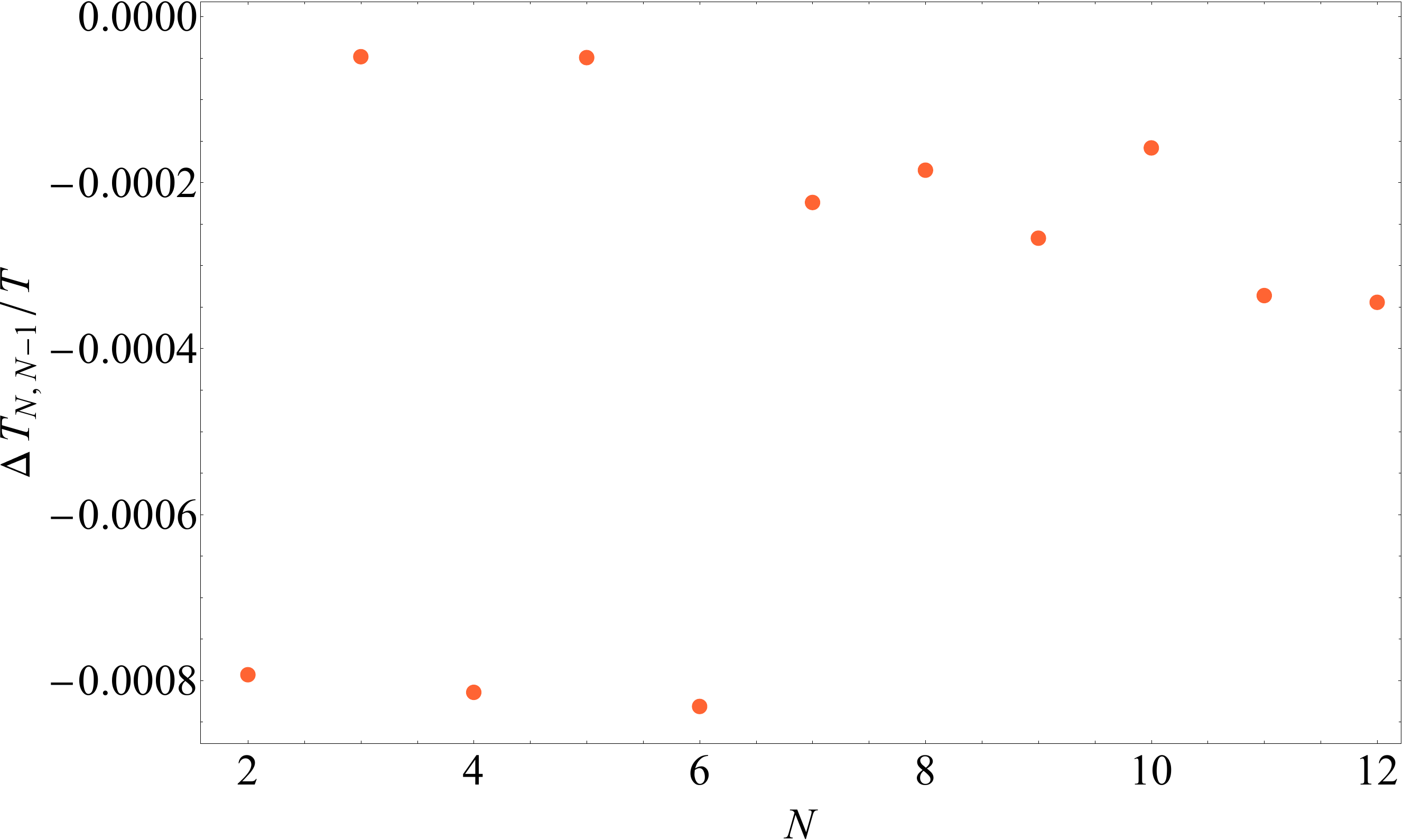}
     \includegraphics[width=0.505\textwidth]{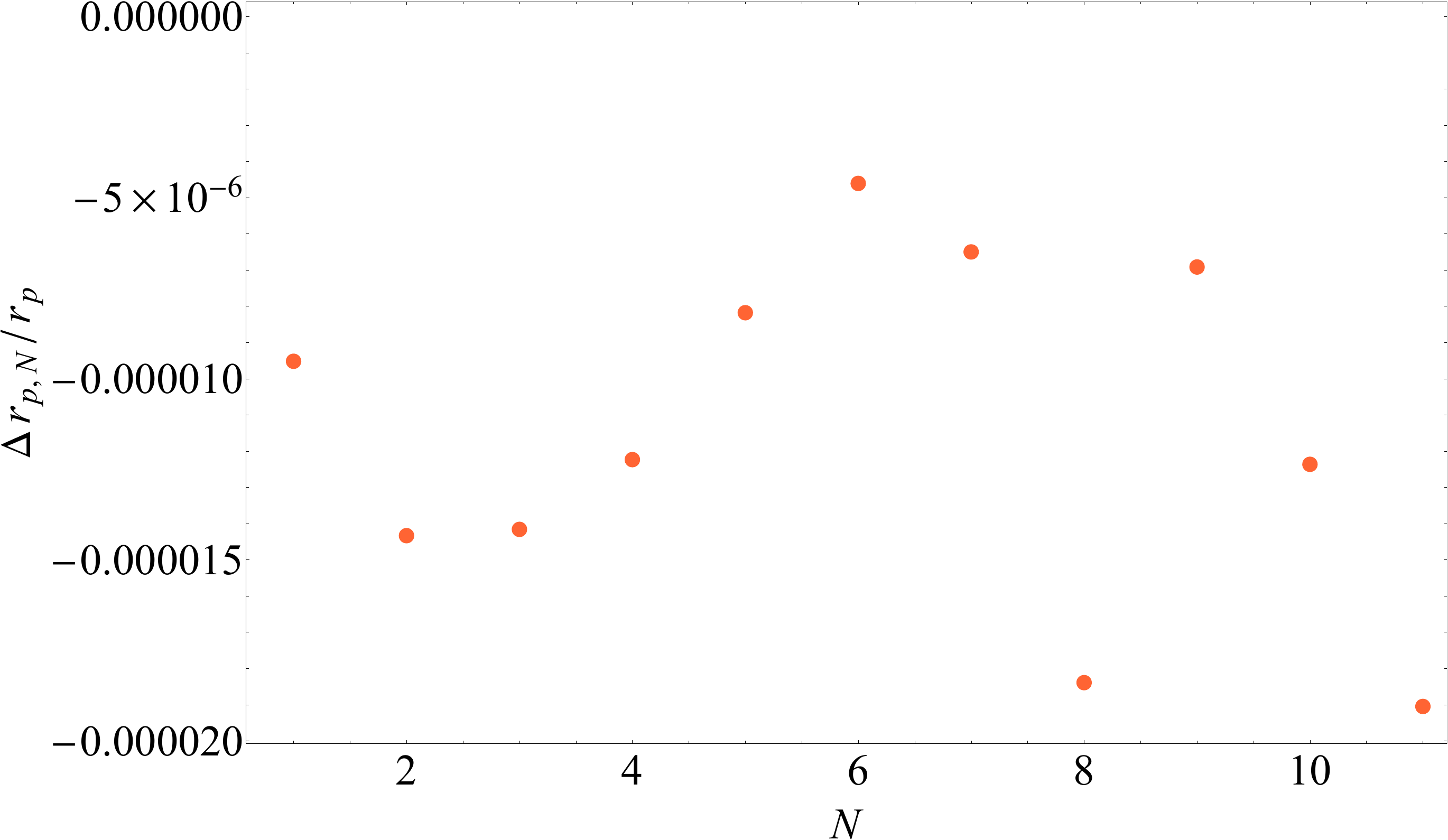}
    \caption{Left: The relative difference of the time between pericenter passages $N$ and $N-1$ to the Keplerian orbital period, e.g., the value appropriate to $N = 2$ represents the time between the first and second pericenter passage, minus and subsequently divided by the Keplerian orbital period $T$. Right: The relative difference between the pericenter distance on the $N^{\rm th}$ pericenter passage to the Keplerian pericenter distance $r_{\rm p}$. The values in these figures are consistent with the analytical expressions derived here; see Equation \eqref{DeltaT}. These data are appropriate to the orange curve in Figure \ref{fig:Omega_of_t}, from which it is apparent that the largest changes in the rotational velocity of the dumbbell correspond to the largest changes in both the orbital period and the pericenter, which is expected.}
    \label{fig:TN_rpN}
\end{figure*}

Figure \ref{fig:Omega_of_t} shows the angular velocity of the dumbbell, $\Omega = d\varphi/dt$, relative to the orbital angular velocity of a Keplerian orbit at pericenter, $\Omega_{\rm p} = \sqrt{GM_{\bullet}\left(1+e\right)}/r_{\rm p}^{3/2}$, as a function of time relative to the Keplerian orbital period, $T = 2\pi a^{3/2}/\sqrt{GM_{\bullet}}$, for $r_{\rm p} = 300 R_{\star}$ and $e = 0.9$. More specifically, the latter two quantities establish the initial conditions for $\dot{r}_{\rm c}$ and $\dot{\phi}_{\rm c}$, which are derived from the Keplerian relations for $r_{\rm c}$ and $\phi_{\rm c}$ (i.e., the initial COM velocities are calculated assuming that the dumbbell is a point mass in the Keplerian potential of the black hole). The initial position is set to $2r_{\rm p}$ and $\phi_{\rm c} = -\pi/2$, so that $\phi_{\rm c} = 0$ coincides with the Keplerian pericenter, and the different curves coincide with the initial angular phases of the dumbbell that are shown in the legend (the initial angular velocity was set to zero). We see that each pericenter passage of the object coincides with a nearly impulsive ``kick'' to its angular velocity, such that the dumbbell is spun up to a phase-dependent fraction of $\Omega_{\rm p}$ on the first pericenter passage, and on each subsequent passage it suffers an effectively instantaneous change to its angular velocity. However, the value of $\Omega/\Omega_{\rm p}$ never exceeds $\sim 2$ over the course of ten orbital periods, demonstrating that the evolution of $\Omega$ is a sequence of sporadic oscillations about $\Omega_{\rm p}$. We also see that the three curves initially display similar evolution owing to the small difference in the initial phase of the dumbbell, but after $\sim 6$ orbits, this similarity is no longer apparent.

The kinetic energy and angular momentum of the dumbbell are small fractions of those appropriate to the COM orbit, and hence the change in the orbital period and pericenter distance of the COM are small. Specifically, since the angular velocity of the dumbbell is on the order of $\Omega_{\rm p} = \sqrt{GM_{\bullet}\left(1+e\right)}/r_{\rm p}^{3/2}$, it follows that the reduction in the orbital period and the pericenter distance are, to leading order in the ratio of the dumbbell radius $R_{\star}$ to the pericenter distance $r_{\rm p}$, 
\begin{equation}
\frac{\Delta T}{T} \simeq \frac{3}{2} \left( \frac{1+e}{1-e}\right)\frac{R_{\star}^2}{r_{\rm p}^2}, \quad \frac{\Delta r_{\rm p}}{r_{\rm p}} \simeq \frac{2 R_{\star}^2}{r_{\rm p}^2}. \label{DeltaT}
\end{equation}
For TDEs on bound orbits with $r_{\rm p} \simeq few\times r_{\rm t} = few\times R_{\star}\left(M_{\bullet}/M_{\star}\right)^{1/3}$, the relative changes in these quantities scale with the mass ratio as $\left(M_{\bullet}/M_{\star}\right)^{-2/3} \ll 1$. The contribution from the factor $(1+e)/(1-e)$ becomes increasingly significant in the limit $e\rightarrow1$. For the specific case considered here, with $e = 0.9$ and $r_{\rm p} = 300R_{\star}$, we have $\Delta T/T \simeq 3.2\times 10^{-4}$ and $\Delta r_{\rm p}/r_{\rm p} \simeq 2.2\times 10^{-5}$. These numbers are in good agreement with Figure \ref{fig:TN_rpN}, the left panel of which shows the relative difference of the time between pericenter passages $N$ and $N-1$ to the Keplerian orbital period $T$, $\Delta T_{\rm N, N-1}$, e.g., the value at $N = 2$ gives the time between the second and first pericenter passage, minus and divided by the Keplerian orbital period. The right panel shows the relative difference between the pericenter distance of the N$^{\rm th}$ pericenter passage and the Keplerian pericenter distance. These data are appropriate to the orange curve in Figure \ref{fig:Omega_of_t}; note that -- in agreement with expectations -- the largest changes in the orbital period and pericenter correspond to the largest angular velocities of the dumbbell.

The expression above for $\Delta T/T$ assumes that the energy imparted to the star is small compared to the binding energy of the COM orbit, but this will only be valid for eccentricities that are not close to unity; indeed, in the limit that the eccentricity is exactly 1, the energy of the new orbit is due exclusively to the energy imparted to the object via tides, and Equation \eqref{DeltaT} cannot be used to calculate the change in the orbital period. By comparing the binding energy of the original orbit to the rotational energy of the star, we see that the above expression is only valid for eccentricities that satisfy

\begin{equation}
    1-e \gtrsim \frac{R_{\star}^2}{\left(1+e\right)r_{\rm p}^2},
\end{equation}
which, for numbers appropriate to typical TDEs, is of the order $1-e \gtrsim 10^{-5}$. When the eccentricity is above this value, a lower bound on the orbital time of the (captured, in this case) object is the object's dynamical time multiplied by the mass ratio of the black hole to the mass of the object, which amounts to thousands of years for typical numbers; see \citet{cufari22b, cufari23} for additional discussion of this case, the latter of which demonstrates that the orbital period is orders of magnitude longer than this lower bound.

This toy problem will overestimate the angular momentum imparted to the star in the limit that $r_{\rm p} \gg r_{\rm t}$, the reason being that the quadrupole moment of the ``dumbbell'' is permanent and maximal (compared to that of a star, which is itself excited by tides). Nonetheless, it is illustrative in demonstrating that a star can only be given an angular velocity comparable to $\sim \Omega_{\rm p}$, implying that -- provided $r_{\rm p}$ is outside the tidal radius -- a star is capable of surviving for many repeated encounters with an SMBH, thus powering the repeated flares observed in ASASSN-14ko. 

Additionally, Equation \eqref{DeltaT} can estimate the period decay rate of rpTDEs. For ASASSN-14ko specifically, a period of $T = 114$ days and an SMBH mass of $10^{7}M_{\odot}$ -- as inferred by \citet{payne23} -- yields a semimajor axis of $a \simeq 1.5\times 10^{15}$ cm. If the pericenter distance is comparable to the tidal radius, then $r_{\rm p} \simeq 2\times r_{\rm t} = 2 R_{\star}\left(M_{\bullet}/M_{\star}\right)^{1/3}$ (we note that high-mass and more evolved stars can give rise to partial TDEs and survive multiple encounters even when the pericenter distance lies closer to the tidal radius; \citealt{coughlin22}), and the eccentricity of the orbit satisfies\footnote{It is interesting to note that the mass of the black hole drops out of this expression.}
\begin{equation}
1-e = \frac{r_{\rm p}}{a} \simeq 2\left(\frac{GM_{\star}T^2}{4\pi^2 R_{\star}^3}\right)^{-1/3} \simeq 0.02\label{eexp}
\end{equation}
Inserting this into Equation \eqref{DeltaT} then gives
\begin{equation}
    \frac{\Delta T}{T} \simeq \frac{3}{2}\left(\frac{1+e}{1-e}\right)\left(\frac{M_{\bullet}}{M_{\star}}\right)^{-2/3} \simeq 0.003.
\end{equation}
This value is in good agreement with the observed period decay rate of ASASSN-14ko (note that this equation is taking the absolute value; the orbital period will be reduced by this amount, owing to the fact that energy is imparted to the star and removed from the COM orbit). However, we recognize that this is the period change following a single encounter, and this energy must be efficiently lost from the system if subsequent encounters are to continually reduce the orbital period by a comparable amount. We return to a discussion of this point in Section \ref{sec:conclusions} below.

There are a number of effects that this simple model does not incorporate -- as we already noted, the quadrupole moment of the ``dumbbell'' considered here is permanent and geometrically maximal, whereas that of a star is produced by the tidal interaction itself and the net rotation of the star is the consequence of a nonlinear coupling between the induced quadrupole and the tidal field. The quadrupole moment of the star will also change as a function of the spin, and the star also has oscillatory modes and loses a fraction of its mass. While one could conceivably model these additional effects through modifications to the geometry and the Lagrangian in Equation \eqref{lagfull}, we instead opt to numerically and hydrodynamically simulate the interaction between a gravitationally self-bound star and an SMBH. We describe our adopted numerical methods, the simulations, and the results in the next section.

\section{Hydrodynamical simulations}
\label{sec:hydro}
\subsection{Simulation Setup}
We used the smoothed particle hydrodynamics (SPH) code {\sc phantom}~\citep{price18} to simulate the repeated partial tidal disruption of stars of different masses and ages by a $10^6 M_{\odot}$ SMBH, where the latter is modeled as a Newtonian point-mass potential. Using the stellar evolution code {\sc mesa}~\citep{paxton11,paxton13,paxton15,paxton18}, we evolved stars from the zero age main sequence (ZAMS), when hydrogen fusion is initiated, to the terminal age main sequence (TAMS), when the hydrogen fraction remaining in the core is $\lesssim 0.1 \%$. The {\sc mesa} profiles were mapped onto a three-dimensional particle distribution in {\sc phantom}, and relaxed using the routine implemented in \cite{golightly19b}, which generates stable density profiles that agree well with the original {\sc mesa} profiles. We used $10^6$ particles to model the stars used in our simulations, which has been shown to be an adequate resolution for partial TDEs in~\cite{miles20,nixon21}. However, we also performed a subset of these simulations at a resolution of $10^7$ particles to test the numerical convergence of our results. Details of the numerical method, such as the implementation of stellar self-gravity and equation of state, are identical to those described in~\citet{coughlin15}. 
\begin{figure}
    \includegraphics[height=5.8cm,width=0.48\textwidth]{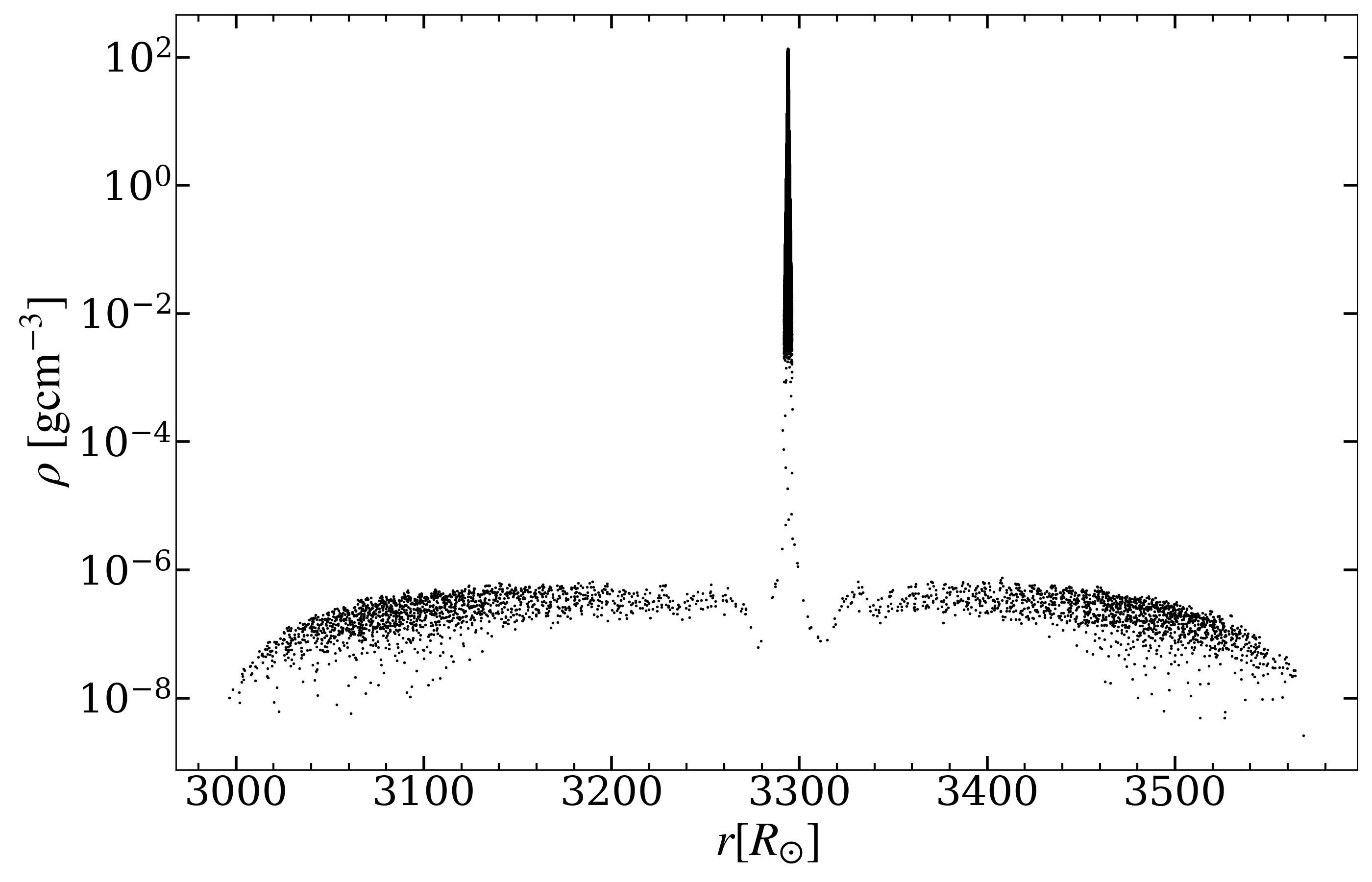}
    \caption{A particle plot showing the density of the disrupted $3 M_{\odot}$ TAMS star as a function of radial distance from the black hole, at $\sim 2$ days after the COM of the star reaches pericenter. The density profile shows a distinct division between particles belonging to the core (densities $\gtrsim 10^{-2}$ g cm$^{-3}$) and those belonging to the tidally disrupted debris stream (densities $\lesssim 10^{-6}$ g cm$^{-3}$). In this case the subset of particles having densities greater than $0.001$ g  cm$^{-3}$ is defined as the core. }
    \label{fig:core-stream-density}
\end{figure}

\begin{figure}
    \includegraphics[height=6.8cm,width=0.48\textwidth]{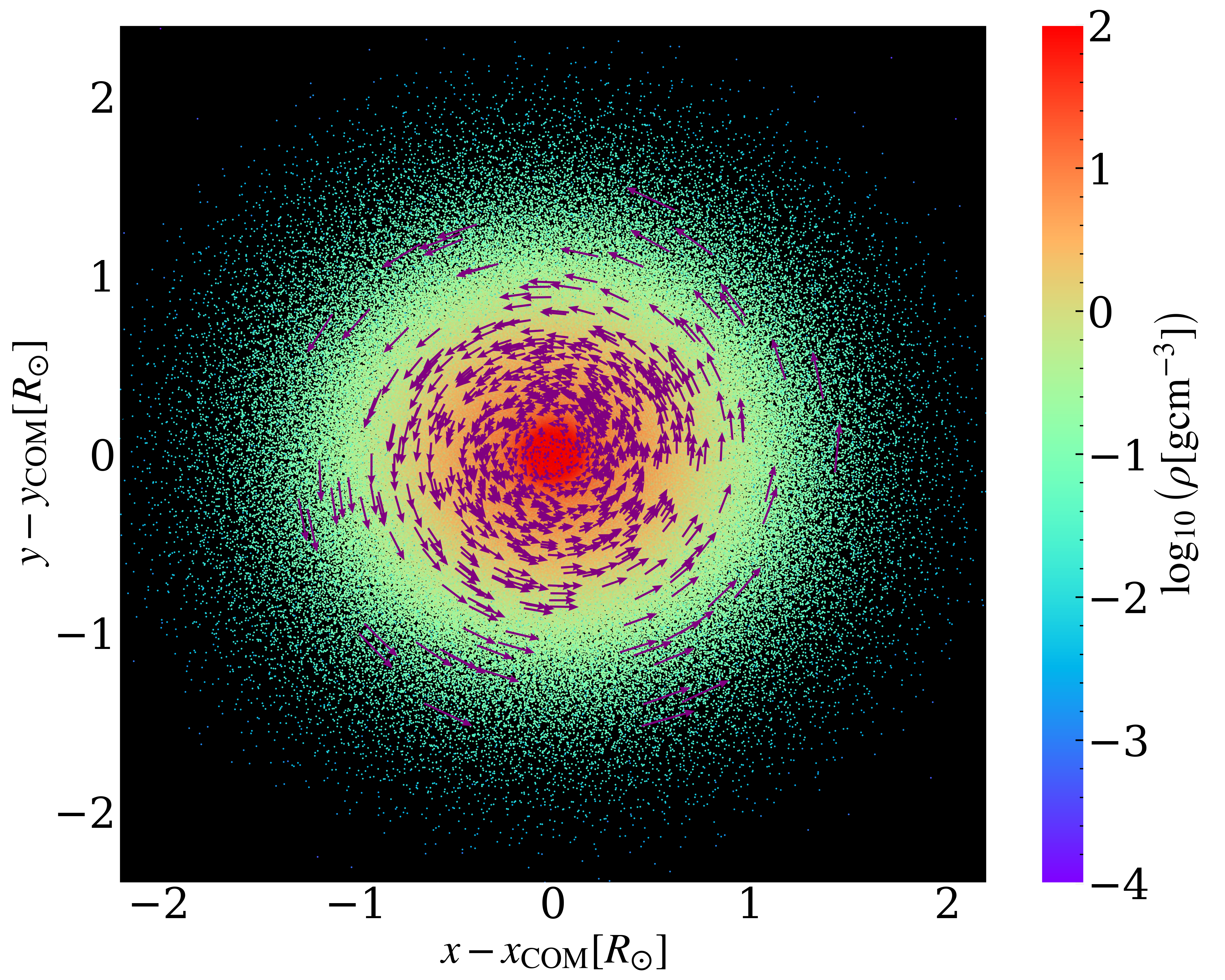}
    \caption{The density of the core in the orbital plane, represented as a function of its COM subtracted position coordinates. The vectors overlaid on the figure represent the velocity of the core particles in the COM frame, with their length indicating the norm of the vectors. The velocity structure shows the spin imparted to the stellar core through tidal interactions. }
    \label{fig:core-velocity}
\end{figure}

To simulate the disruption process, the relaxed star was placed at an initial distance of $5 r_{\rm t}$ from the SMBH, with the COM on a parabolic Keplerian orbit. Upon passing through pericenter, the tidal field of the SMBH partially disrupts the star and strips off a fraction of its outer envelope. Approximately half of the stripped debris is gravitationally bound to the SMBH and accretes onto it. The surviving core in a partial TDE reduces the hydrodynamical timestep of the simulations relative to the dynamical time of the disrupted debris. Thus, to calculate the fallback rate of bound stellar debris, the surviving stellar core was replaced with a point mass roughly 2 days after the COM reached pericenter, at which point the core had receded to a distance of  $>15 r_{\rm t}$ from the SMBH (for low mass stars, having smaller tidal radii, this distance is larger) as was done in, e.g.,~\cite{miles20,golightly19b}. The fallback rate is calculated as the rate at which particles from the disrupted debris stream return to the accretion radius of the SMBH, which is defined as the inner $3 r_{\rm t}$ of the computational domain for our simulations. The fallback rate of particles from the disrupted debris stream closely tracks the accretion rate onto the SMBH, subject to the assumption of efficient disc circularization and negligible viscous delays (see \citealt{mockler19,nicholl22} for observational evidence that suggests that viscous delays are small).

Simulating the subsequent encounters by evolving the star on its highly eccentric orbit is computationally intractable, since the orbital time of these events, which ranges from months-to-years, amounts to thousands of dynamical times of the disrupted star. To surmount this issue, we simulated each encounter by allowing the COM of the star to evolve on its parabolic orbit around the SMBH for $\sim 2$ days past its pericenter, which amounts to $\gtrsim 100$ dynamical times of the star (but still orders of magnitude shorter than the orbital time). The surviving core was then translated back to the initial position of the original star (equal to $5 r_{\rm t}$) by calculating and then subtracting the COM velocity and position from every SPH particle, and subsequently adding back (to each particle in the core) those of the original stellar orbit. Figure~\ref{fig:core-stream-density} shows the density profile of the partially disrupted star for a $3 M_{\odot}$ TAMS star, at a time when the COM has evolved for $\sim 2$ days past its pericenter. As seen in the figure, there is a distinct dichotomy between the core and stream particles, with the exception of a few particles that are situated in the intermediate region. For each simulation, we chose a density cutoff that lies between the tidally disrupted debris stream and the dense central structure that we identify as the core. The core, comprised of particles having a density greater than the threshold value, was translated back on its orbit to simulate the subsequent encounters. We verified that once a distinct core had formed, the density profile remained effectively unmodified, implying that performing the core-translation routine a day earlier (or later) does not have a measurable effect on the fallback rate. The fallback rate for each successive orbit was obtained by replacing the core with a sink particle using the procedure described above, and tracking the rate of return of particles from the debris stream to the SMBH.

\begin{figure*}
    \includegraphics[height=11cm, width=0.995\textwidth]{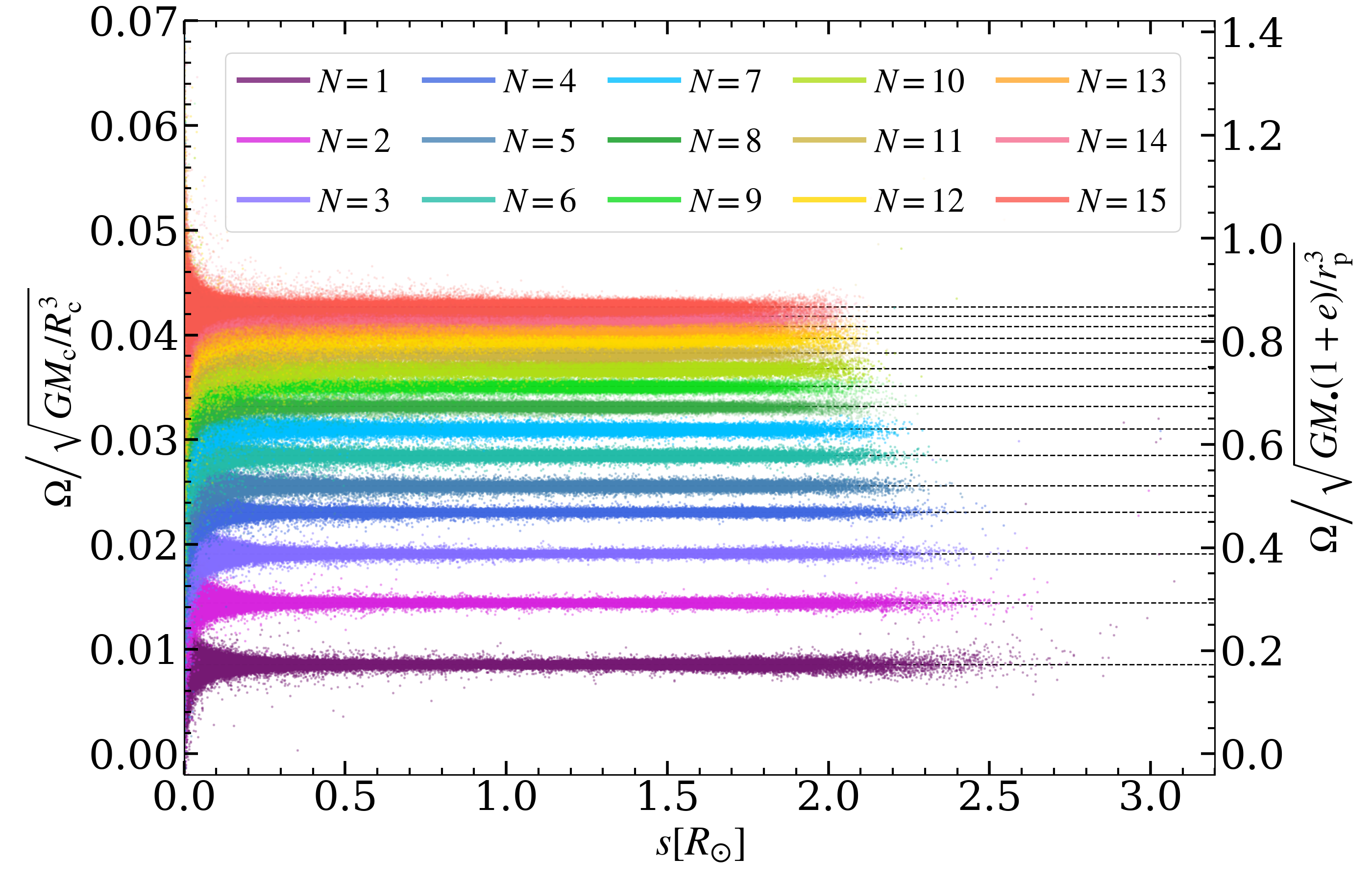}
    \centering
     \includegraphics[height=11cm, width=0.9\textwidth]{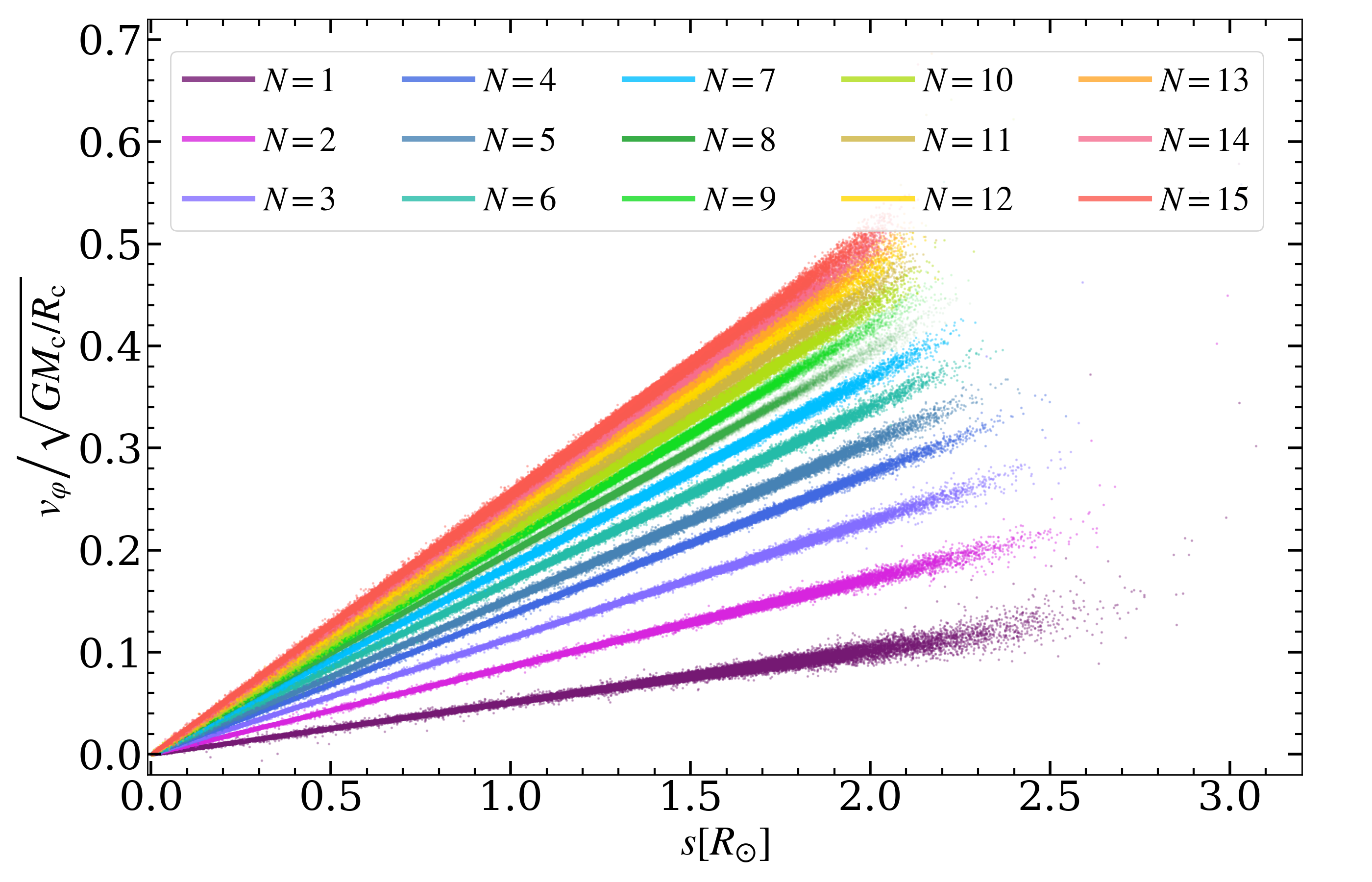}
    \caption{Results of the SPH simulation for the $3 M_{\odot}$ TAMS star. The top panel shows the angular velocity $\Omega$ imparted to the core, with the primary $y-$axis normalized by the break-up angular velocity of the core, $\sqrt{G M_{\rm c}}/R_{\rm c}^{3/2}$, and the secondary $y-$ axis normalized by the angular velocity of the COM at pericenter. The bottom panel shows its linear component, $v_{\varphi} = s \Omega $ in units of $\sqrt{G M_{\rm c}/R_{\rm c}}$, as a function of cylindrical distance $s$ from the center. }
    \label{fig:angularmomentum}
\end{figure*}

\subsection{Results for a \texorpdfstring{$3M_{\odot}$}{Lg} TAMS star}
\label{sec:hydro3M}
One of our primary goals in this work is to test the viability of the rpTDE model in providing an explanation for the repeated flares observed from ASASSN-14ko. To do this, we describe here the results of simulating the repeated partial disruption of a $3 M_{\odot}$ TAMS star by a $10^6 M_{\odot}$ SMBH, with an impact parameter $\beta=1$. The impact paramater $\beta \equiv r_{\rm t} / r_{\rm p}$ quantifies the strength of the encounter between the star and the SMBH. For a high-mass star at its late evolutionary stages, the analytical prediction for the critical impact parameter $\beta_{\rm c}$ required for its complete disruption exceeds this value by an appreciable amount (\citealt{coughlin22}; from Figure~4 of~\cite{bandopadhyay24} a $3M_{\odot}$ TAMS star has a critical impact parameter $\beta_{\rm c} \sim 6$, though we note that general relativistic effects may reduce this value somewhat for very massive SMBHs; e.g., \citealt{beloborodov92, kesden12, gafton15, tejeda17, gafton19, ryu20, jankovic23}). Thus, $\beta=1$ results in only a small amount of mass lost, and the star should -- subject to the influence of the imparted rotation discussed above -- be able to survive for many encounters while simultaneously fueling accretion events. 

Each successive pericenter passage spins up the tidally stripped core  by a small fraction of its break-up velocity. Figure~\ref{fig:core-velocity} shows the velocity of the fluid comprising the core in its COM coordinates (i.e., both COM position and velocity removed), which qualitatively demonstrates that the core is indeed rotating following its tidal interaction with the SMBH. In the top panel of Figure~\ref{fig:angularmomentum}, we show the angular velocity $\Omega$ imparted to the core during the first fifteen pericenter passages as a function of cylindrical radius $s$ within the core. The primary $y-$axis shows the angular velocity normalized by the break-up angular velocity of the core, $\Omega_{\rm c} = \sqrt{G M_{\rm c}/R_{\rm c}^3}$. Here, $R_{\rm c} \approx 0.2 R_{\odot}$ is the core radius for a $3M_{\odot}$ TAMS star, defined as the radial distance from the center of the star at which the self-gravitational field of the star is maximized \citep{coughlin22}, and $M_{\rm c} \approx 0.3 M_{\odot}$ is the mass contained within the core. The secondary $y-$axis shows the angular velocity normalized by the angular velocity of the COM at pericenter, $\Omega_{\rm p}=\sqrt{2G M_{\bullet}/r_{\rm p}^3}$. This figure demonstrates that while the rotation rate is comparable to $\Omega_{\rm p}$, the imparted angular velocity is a small fraction of the break-up angular velocity of the core. The core is thus relatively unmodified, and this allows the bulk of the star to survive many encounters as it tidally interacts with the SMBH. The bottom panel of Figure~\ref{fig:angularmomentum} shows the corresponding linear velocity $v_{\varphi} \equiv s \Omega$, normalized by $\sqrt{G M_{\rm c}/R_{\rm c}}$. 

The secular increase in the imparted angular velocity $\Omega$ arises from the fact that, unlike the ``dumbbell'' analyzed in the previous section that has a permanent quadrupole moment, the quadrupole moment of the star is raised and subsequently torqued by the tidal interaction with the SMBH during its pericenter passage. The imparted angular velocity is therefore a nonlinear response in the ratio of the tidal force to the self-gravitational force, and also depends on the angle that the tidal bulge makes with the line joining the star to the SMBH in the corotating frame of the star; some differences also likely arise in part from the fact that the size of the star -- and hence the size of the lever arm that mediates the torque applied by the SMBH -- increases as a result of the imparted stellar rotation and the structural changes in response to the mass loss. The oscillatory modes of the star are also damped on a viscous time, which in our simulations is numerical and small but still shorter than the time before which the star is translated back to pericenter; if the oscillatory modes persisted and were not damped, there would also be a pseudo-random oscillation in the angular momentum, akin to what is seen in Figure \ref{fig:Omega_of_t}. We discuss the nature of the imparted rotation further in Section \ref{sec:spinup}.

Additionally, we note that in our simulations, the thermal energy that is dissipated viscously (numerically) is not put back into the gas, and is instead assumed to be efficiently lost from the system; this choice is motivated by the fact that the outer layers of the star can be subject to excessive numerical heating (see, e.g., \citealt{norman21, coughlin22b}). There should, however, be some level of physical heating in response to the dissipation of kinetic energy imparted by tides, and depending on where the thermal energy is deposited, it may instead be trapped within the star, leading to its more rapid destruction. To assess the impact of viscous heating and to demonstrate that it does not have a strong impact on our results and corresponding conclusions, in Appendix \ref{sec:tidal heating} we present a detailed investigation of the effects of both numerical resolution and the thermodynamic prescription.

\begin{figure*}
    \includegraphics[height=6cm,width=0.485\textwidth]{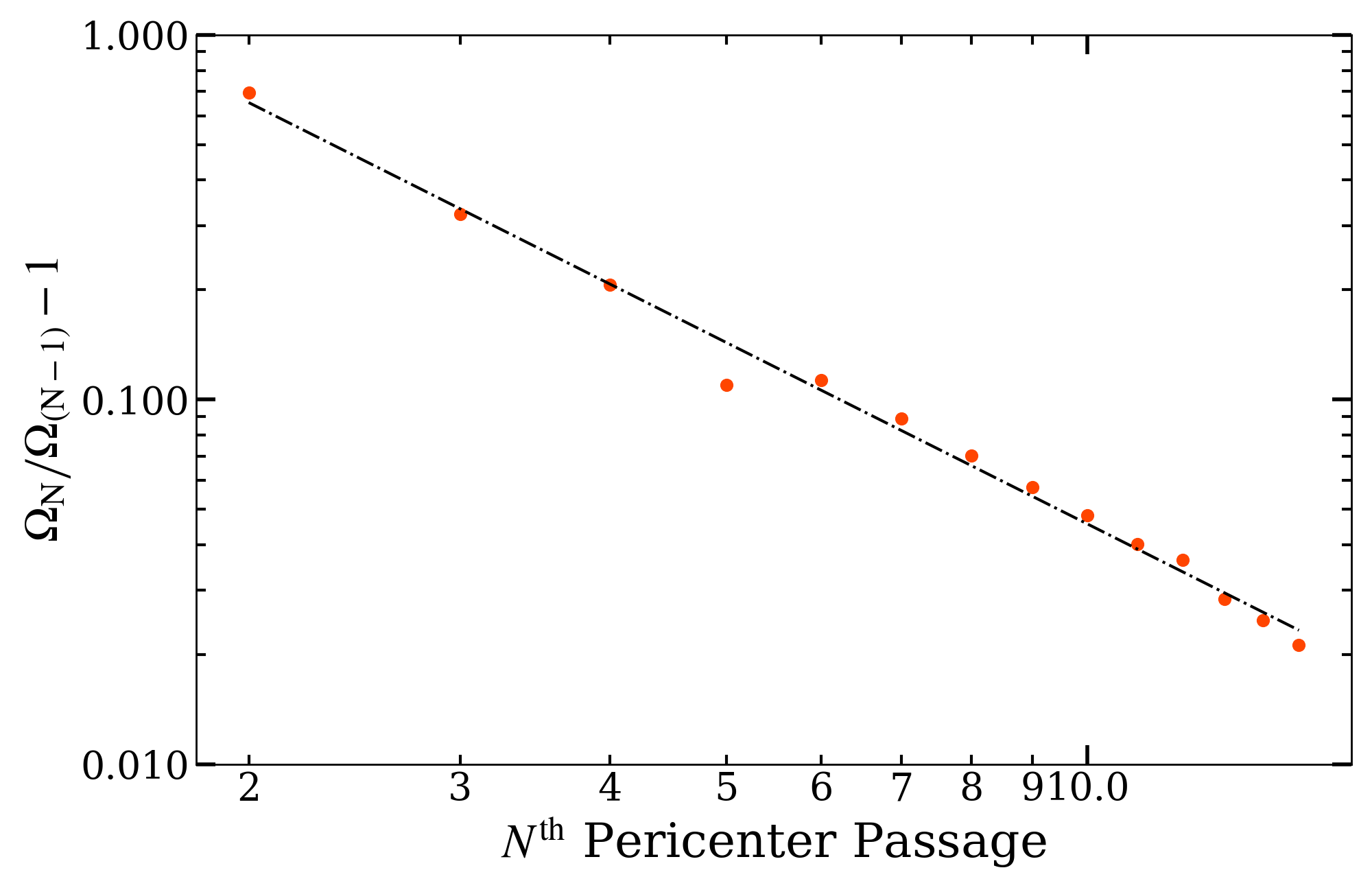} 
     \includegraphics[height=6.1cm,width=0.485\textwidth]{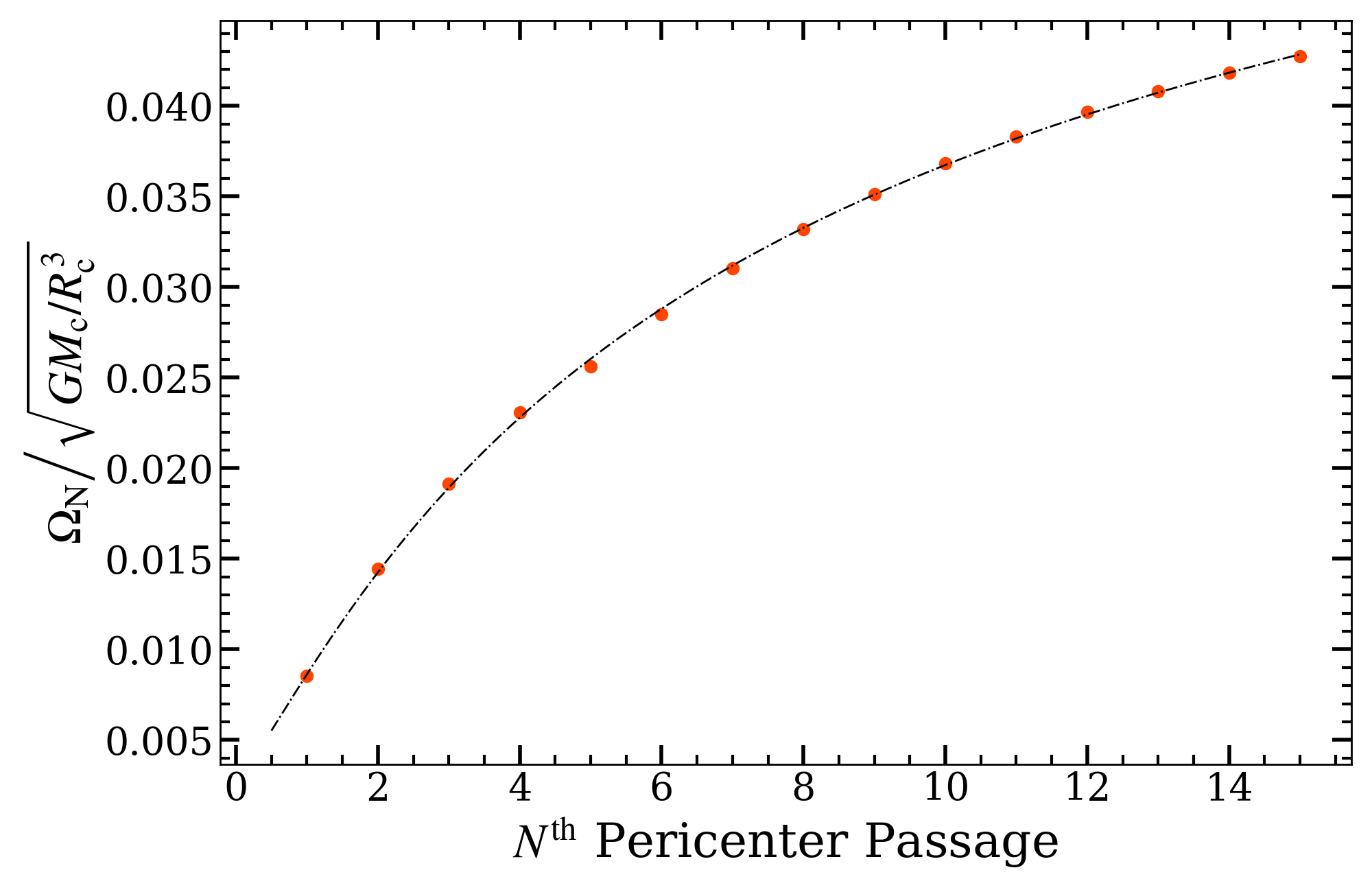}
    \caption{Left: The fractional change in the angular velocity imparted to the core between the $N^{\rm th}$ and $(N-1)^{\rm th}$ pericenter passages. The downward trend indicates that the increment in angular velocity between consequtive pericenter passages decreases with an increase in the number of pericenter passages, $N$. Right: The angular velocity imparted to the core, in units of its break-up spin $\sqrt{G M_{\rm c}/R_{\rm c}^3}$, as a function of the number of pericenter passages.  }
    \label{fig:omegavsN}
\end{figure*}

The increment in the angular velocity imparted to the core between successive pericenter passages decreases as the number of pericenter passages, $N$, increases, and it converges to a constant value in the large-$N$ limit. To illustrate the convergence of the angular velocity imparted to the core, in the left panel of Figure~\ref{fig:omegavsN} we fit the fractional change in $\Omega$ between successive pericenter passages to a power law in $N$. The power-law index obtained from the fit is $\sim -1.65$, thus giving $\frac{1}{\Omega}\frac{d\Omega}{dN} \propto N^{-1.65}$. Using this, we then solve for $\Omega(N)$, which is shown in the right-hand panel of Figure~\ref{fig:omegavsN}. The asymptotic limit of this function is $\sim 0.07$, thus showing that the imparted angular velocity $\Omega$ converges to a value of $\sim 0.07 \Omega_{\rm c}$ in the limit that $N\rightarrow \infty$.

\begin{figure}
    \includegraphics[width=0.48\textwidth]{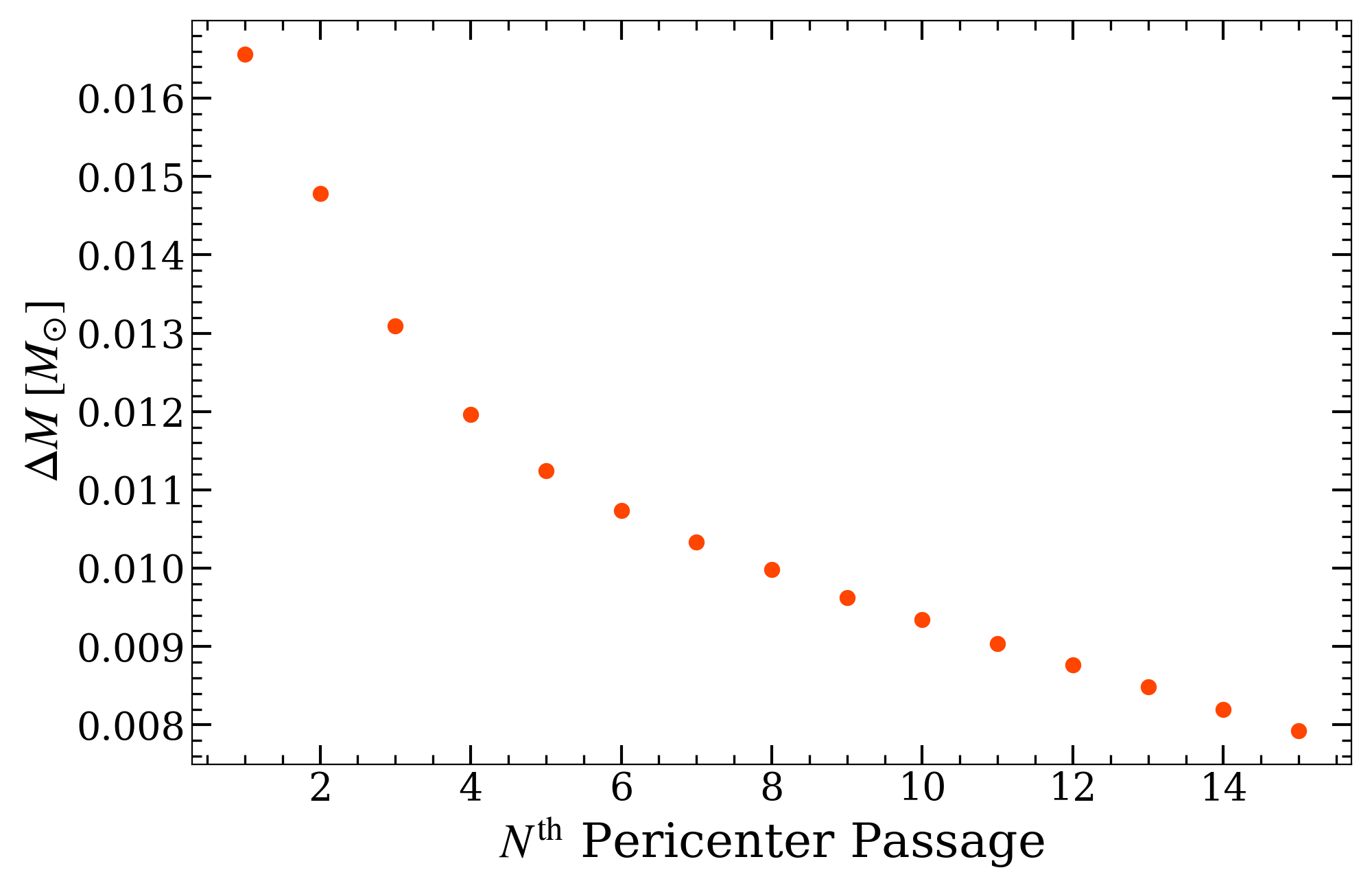}
     \caption{The amount of mass stripped $\Delta M$, as a function of number of pericenter passages $N$, for the $3 M_{\odot}$ TAMS star on an orbit having an impact parameter $\beta=1$.}
    \label{fig:mass_stripped_vsN}
\end{figure}

\begin{figure*}
    \centering
    \includegraphics[height=11cm,width=0.995\textwidth]{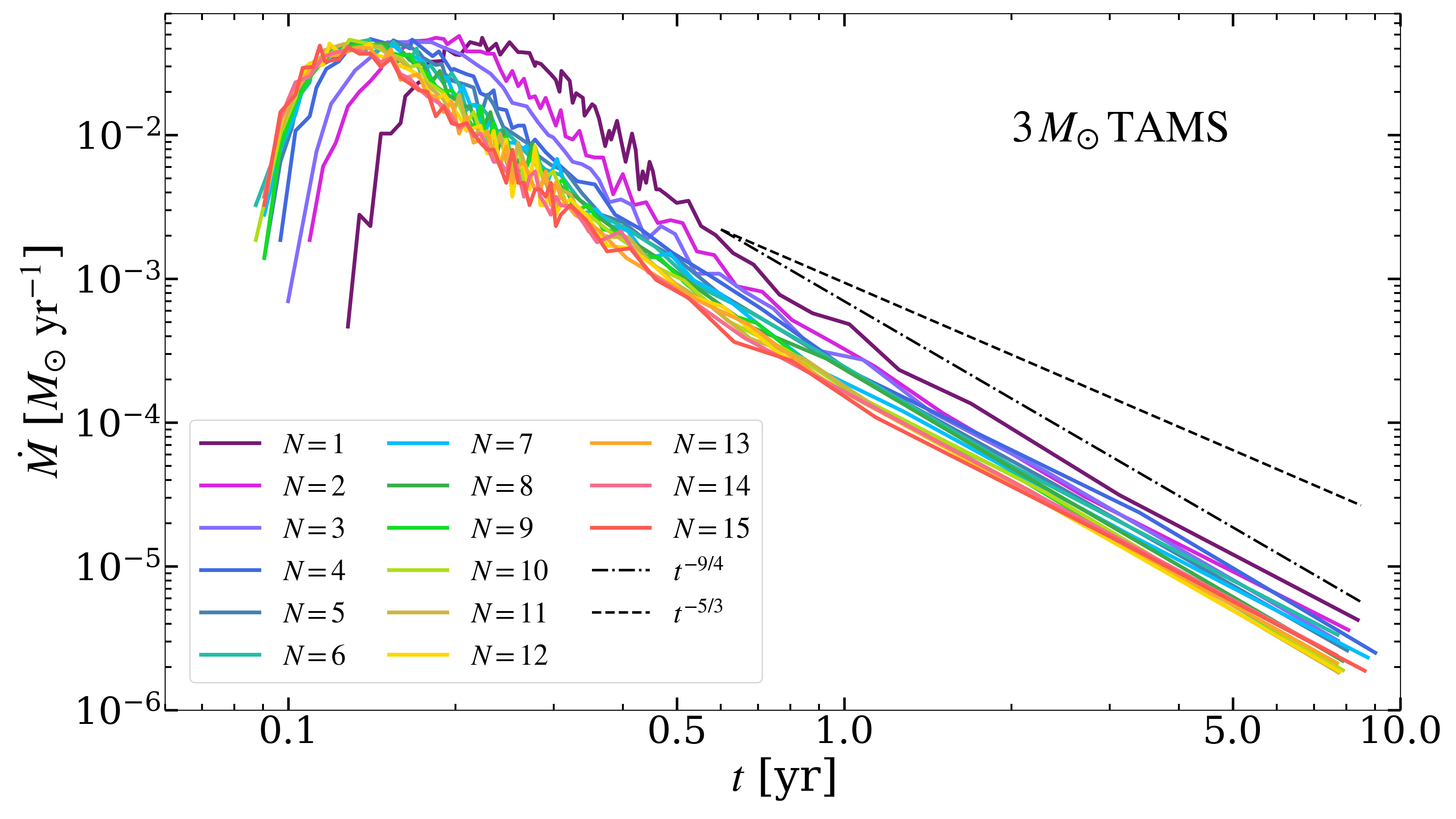}
    \caption{Fallback rates from a simulation of the repeated partial disruption of a $3 M_{\odot}$ TAMS star by a $10^6 M_{\odot}$ SMBH, with a resolution of $10^6$ particles for the disrupted star. Each encounter is labelled by a different color, as indicated by the number in the legend. The impact parameter for each encounter was $\beta = 1$. Roughly the same amount of mass is stripped off from the star during each subsequent pericenter passage, thus giving rise to a comparable peak luminosity between pericenter passages. The late-time evolution of the fallback rates scales as $t^{-9/4}$, consistent with the expected scaling for partial disruptions. }
    \label{fig:fallbackrates}
\end{figure*}

\begin{figure}[h]
   
    \includegraphics[height=6cm,width=0.48\textwidth]{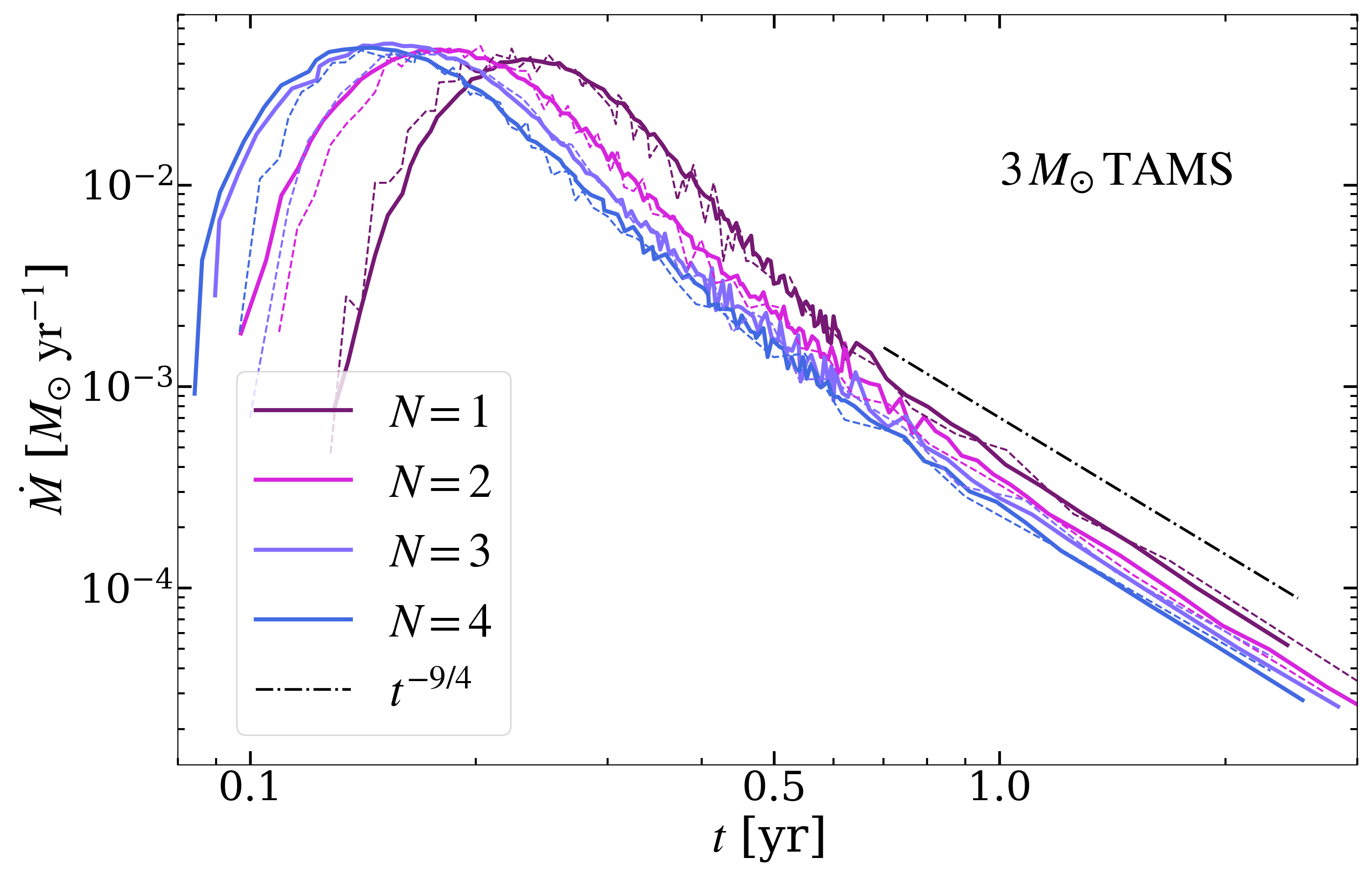}
     \caption{Comparison of fallback rates at different resolutions for the $3M_{\odot}$ TAMS star. The solid curves have a higher resolution of $10^7$ particles, and the dashed curves represent the lower resolution simulations with $10^6$ particles. The curves for corresponding encounters are in good agreement with each other, barring slight differences in the early-time fallback, which is sensitive to the resolution.}
    \label{fig:fbr_convergence}
\end{figure}

The amount of mass stripped during each encounter is $\mathcal{O} \left(0.01 M_{\odot}\right)$, and gradually decreases over time, as shown in Figure~\ref{fig:mass_stripped_vsN}. The fallback rates of stellar debris\footnote{In our simulations the star is approximated to be on a parabolic orbit, in which case only half of the mass that is tidally stripped from the star returns to the SMBH, with the other half unbound. Because the star must actually be on an elliptical orbit, a fraction of the other (usually unbound) tail can be bound to the SMBH, and will thereby produce a second, relatively low-level (but potentially observable) accretion event; the amount of mass in this ``less bound'' tail depends on the eccentricity of the orbit, the impact parameter $\beta$, and the SMBH mass \citep{hayasaki18, park20, cufari22a}.} onto the SMBH for the first fifteen encounters are shown in Figure~\ref{fig:fallbackrates}. The fallback rates are calculated using the binning procedure described in, e.g.,~\cite{golightly19b,miles20}: at early times and when the particle flux is relatively large we calculate the fallback rate using equal time bins, whereas the late time fallback rate is binned by particle number. All of the fallback rates follow the late-time temporal scaling of $t^{-9/4}$, conforming to the expected scaling for partials~\citep{coughlin19}. Owing to the small amount of mass stripped in these encounters, the fallback rates obtained from the $10^6$ particle simulations exhibit noisy behavior.
To vet the accuracy of the fallback rates obtained, we performed higher-resolution simulations using $10^7$ particles for the first four encounters. The fallback rates for the high-resolution simulations are compared against their low resolution counterparts in Figure~\ref{fig:fbr_convergence}. In general, we find good agreement between the fallback rates for corresponding encounters at different resolutions, with the most noticeable differences arising in the return time of the most-bound debris; because the outer layers of the star are better resolved at higher resolution, and the outer layers of the star coincide with the most-bound debris, the high-resolution simulations generally display earlier return times. The peak timescale decreases gradually from $\sim 80$ days to $\sim 45$ days over the first few encounters, and settles to an almost constant value of $\sim 45$ days beyond the first five encounters. The magnitude of the peak of the fallback rates is nearly constant, at $\sim 0.04 M_{\odot} \rm yr^{-1}$, over all of the fifteen encounters shown in Figure~\ref{fig:fallbackrates}.

\begin{figure}
    \includegraphics[height=6cm,width=0.48\textwidth]{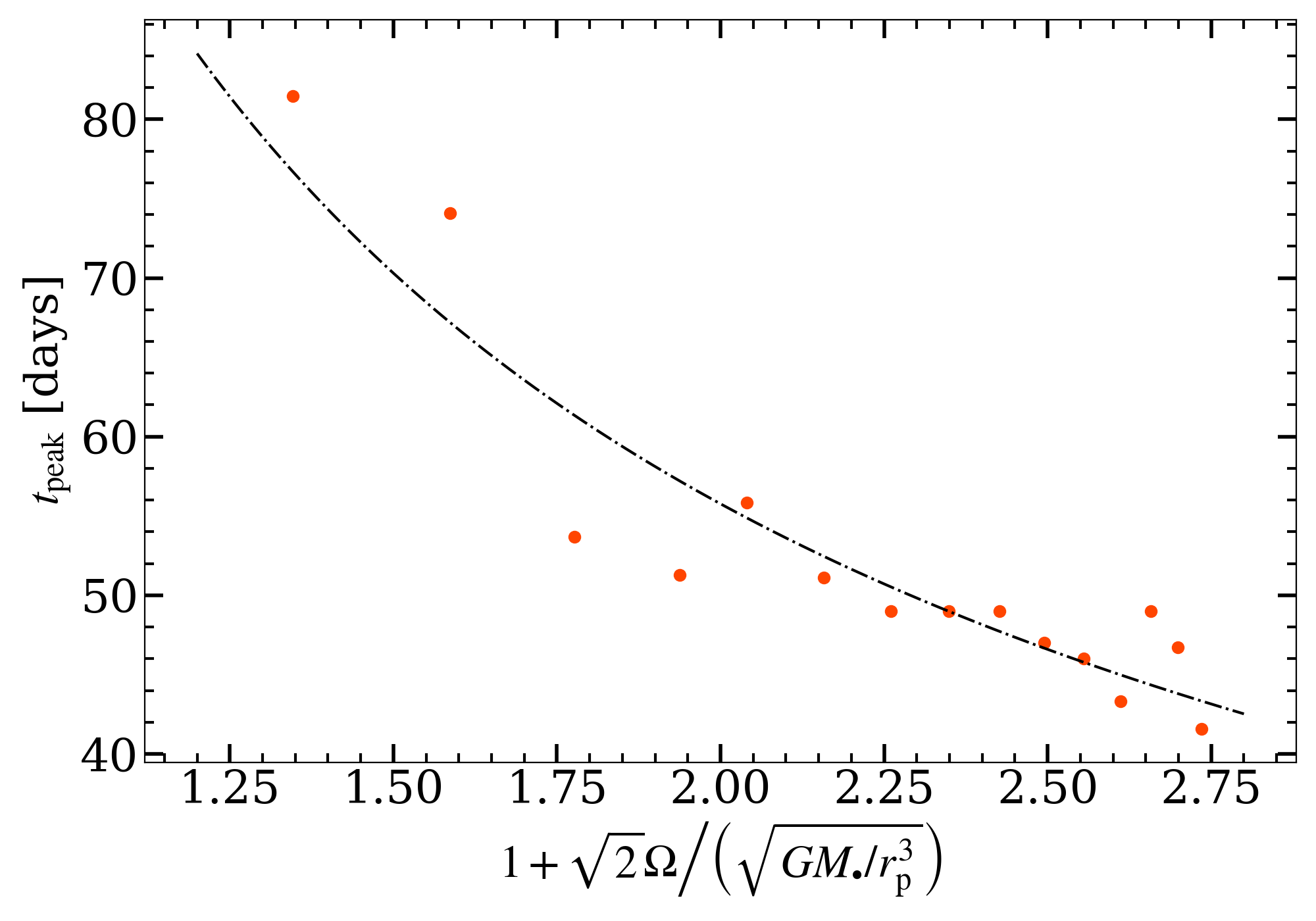}
     \caption{The timescale on which the fallback rates reach their peak, as a function of the imparted angular velocity $\Omega$. The peak timescale declines as a function of $\left( 1 + \sqrt{2} \Omega/\sqrt{G M_{\bullet}/r_{\rm p}^3} \right)$, with a power-law index of $-0.8$.}
    \label{fig:tmb_vs_omega}
\end{figure}

The angular momentum imparted to the core also impacts the energy spread of the tidally disrupted debris. Treating the rotational energy of the star as a perturbation to the Keplerian energy of its orbit, \cite{golightly19a} showed that, for a star rotating in a prograde sense with its axis of rotation aligned with the orbital angular momentum, the leading-order correction to the energy-period relation gives the following expression for the return time of the most bound debris:
\begin{equation}
    \label{eq:tpeak-omega}
    t_{\rm mb} = \left( \frac{R_{\star}}{2}\right)^{3/2} \frac{2 \pi M_{\bullet}}{M_{\star} \sqrt{G M_{\bullet}}}\left( 1+ \sqrt{2} \lambda \right)^{-3/2}.
\end{equation}
In the above expression, $M_{\bullet}$ is the mass of the SMBH, $M_{\star}$ and $R_{\star}$ are the mass and radius of the star, and $\lambda = \Omega/\Omega_{\rm p}$ is the angular velocity of the star normalized by the angular velocity of the COM at pericenter. Since $t_{\rm mb} \propto \left( 1+ \sqrt{2} \lambda \right)^{-3/2}$, we expect the peak fallback timescale to shift to earlier times as the stellar rotation rate increases. In Figure~\ref{fig:tmb_vs_omega}, we plot the peak timescale for the fallback rates against $\left( 1 + \sqrt{2} \lambda \right)$ for the first 15 encounters. As seen in the figure, the time taken for the fallback rates to peak decreases as the rotational velocity imparted to the core increases, and the best-fit line for $t_{\rm peak}$ has a power-law dependence on $\left( 1 + \sqrt{2} \lambda \right)$ with a power-law index of $-0.8$. While we thus recover the general trend that increasing $\Omega$ yields a shorter peak fallback time, the power-law index of $-0.8$ is discrepant with the prediction of~\cite{golightly19a}. This can be attributed to the fact that for the partial disruptions considered here, the surviving core plays an important role in modifying the energy spread of the disrupted debris, and thus the frozen-in approximation -- which is invoked in the derivation of Equation~\eqref{eq:tpeak-omega} -- is not valid. Additionally, the prediction by \citet{golightly19a} is for the return time of the most-bound debris, and the trend may not be as strong for the peak in the fallback rate.

\subsection{Other Stars}
\subsubsection{\texorpdfstring{$1.3 M_{\odot }$}{Lg} TAMS star}
\begin{figure}
    \includegraphics[width=0.48\textwidth]{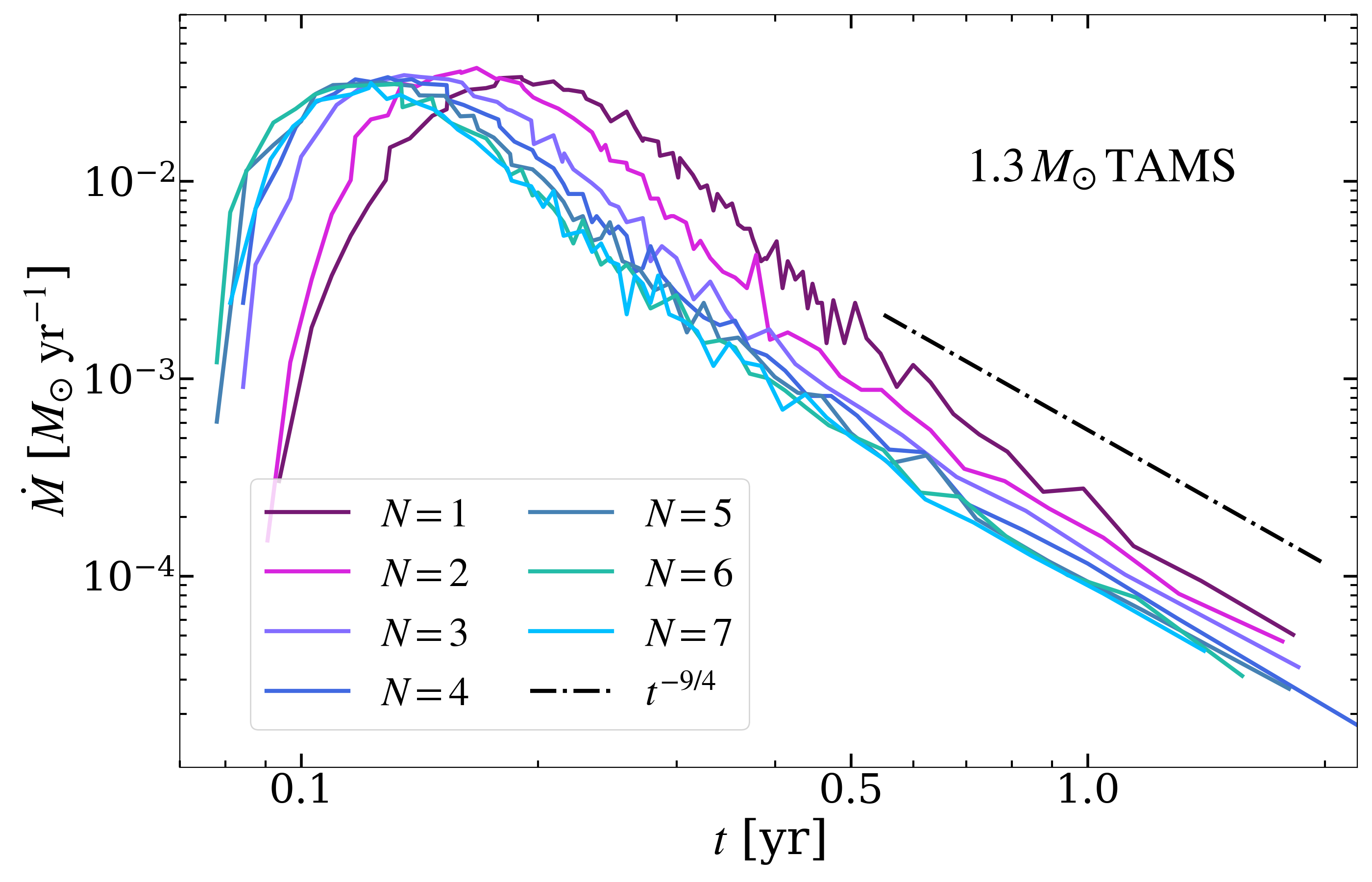}
     \caption{Fallback rates for the first seven encounters between a $1.3M_{\odot}$ TAMS star and a $10^6 M_{\odot}$ SMBH, with an impact parameter $\beta=1$. We used a resolution of $10^6$ particles for these simulations. The evolution of the fallback rates over multiple encounters follows a similar trend to the $3M_{\odot}$ TAMS star.}
    \label{fig:fbr_1p3msunTAMS}
\end{figure}
The $1.3 M_{\odot}$ TAMS star has a density profile similar to the $3 M_{\odot}$ star at the same evolutionary stage. Massive stars such as these, in their late evolutionary stages, develop a core-envelope structure, with a high density core surrounded by a low density and tenuous envelope (see Figure~9 of \citealt{golightly19b} for a comparison of density profiles of low-mass and high-mass stars at different ages of their main sequence). The high-density core makes it increasingly difficult to strip off mass, leading to a progressively declining amount of mass being lost with an increase in the number of pericenter passages. Figure~\ref{fig:fbr_1p3msunTAMS} shows the fallback rates for the first seven pericenter passages of the $1.3 M_{\odot}$ TAMS star orbiting a $10^6 M_{\odot}$ SMBH with an impact parameter $\beta=1$. The evolution of the fallback rates follows a similar trend to the $3 M_{\odot}$ star. As the surviving stellar core is spun up, the peak timescale shifts to earlier times~\citep{golightly19a}, and we see that the magnitude of the peak fallback rate remains almost constant over multiple encounters. These results demonstrate that a high-mass star at its late evolutionary stages, on a grazing orbit around an SMBH, can survive multiple encounters, generating ASASSN-14ko-like flares over a prolonged period.

\subsubsection{\texorpdfstring{$1 M_{\odot }$}{Lg} ZAMS star}
\begin{figure}
    \includegraphics[width=0.48\textwidth]{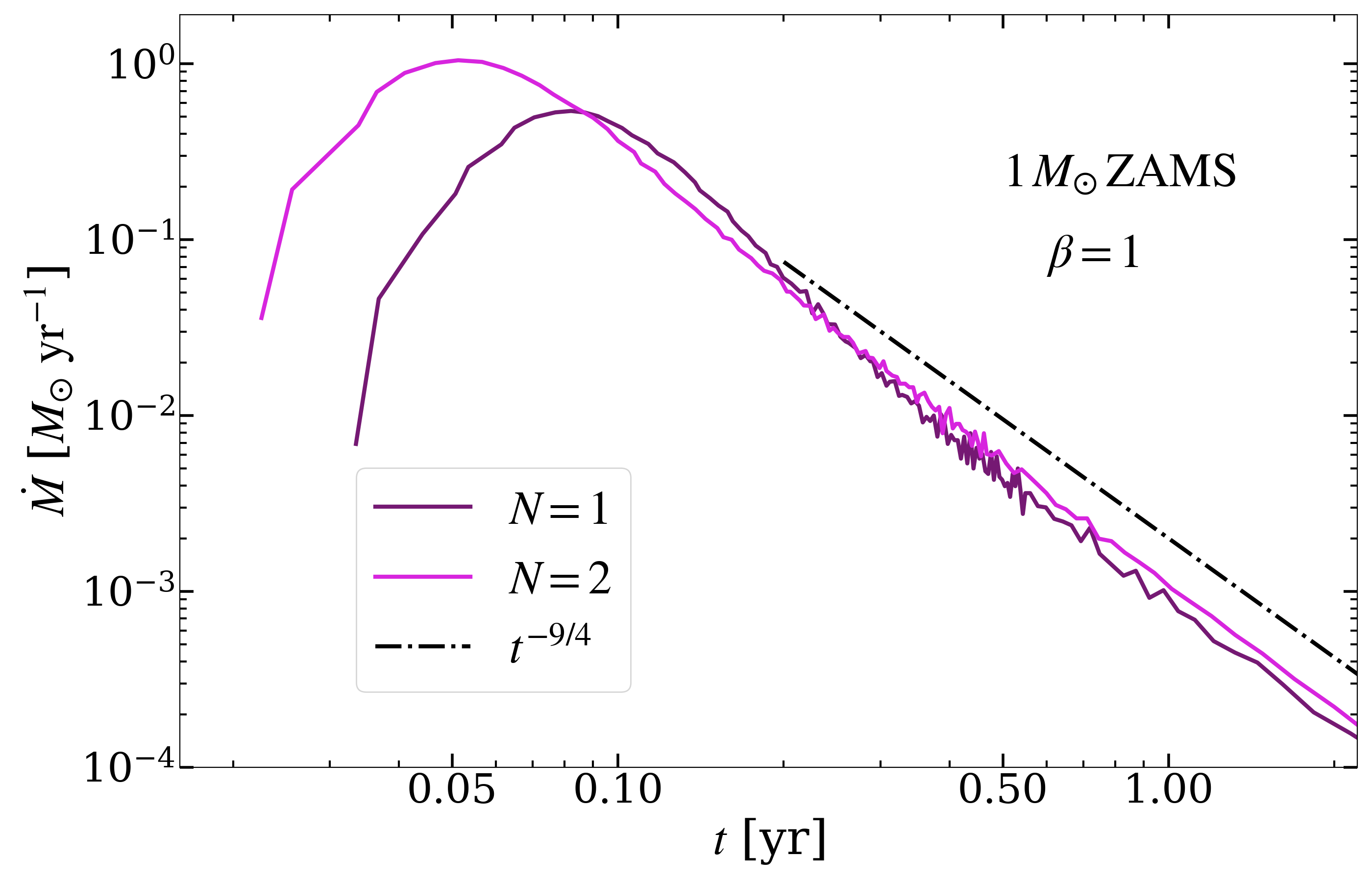}
     \caption{Fallback rates for the disruption of the $1 M_{\odot}$ ZAMS star at $\beta=1$. The second encounter yields a brighter peak relative to the first one, as the amount of mass stripped is roughly doubled. }
    \label{fig:fbr-1msun-beta1}
\end{figure}
Figure~\ref{fig:fbr-1msun-beta1} shows the fallback rates for the first two encounters of a $1 M_{\odot}$ ZAMS star at $\beta=1$. The mass stripped on the first and second pericenter passages is $\sim 0.1 M_{\odot}$ and $\sim 0.2 M_{\odot}$ respectively, yielding a much brighter peak for the second encounter, as seen in the figure. Clearly, a $1 M_{\odot}$ ZAMS star on an orbit having an impact parameter $\beta=1$ cannot survive more than a few encounters, and it thus cannot reproduce the observed lightcurve for ASASSN-14ko. However, the fallback rates are consistent with the observed lightcurve of AT2020vdq, for which two flares, with the second one having a brighter peak, have currently been observed~\citep{somalwar23}.

\begin{figure}
    \includegraphics[width=0.48\textwidth]{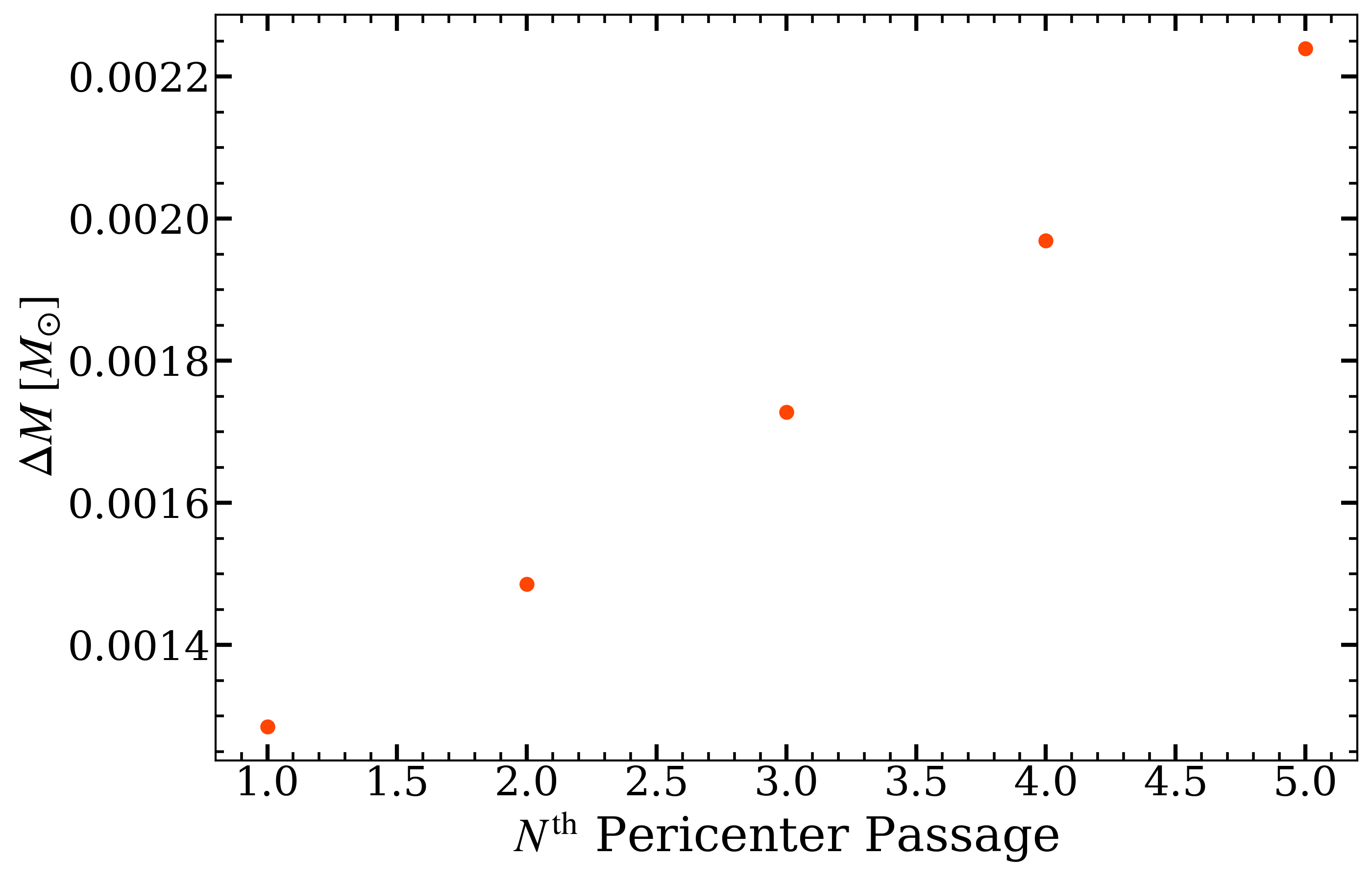}
    \includegraphics[width=0.48\textwidth]{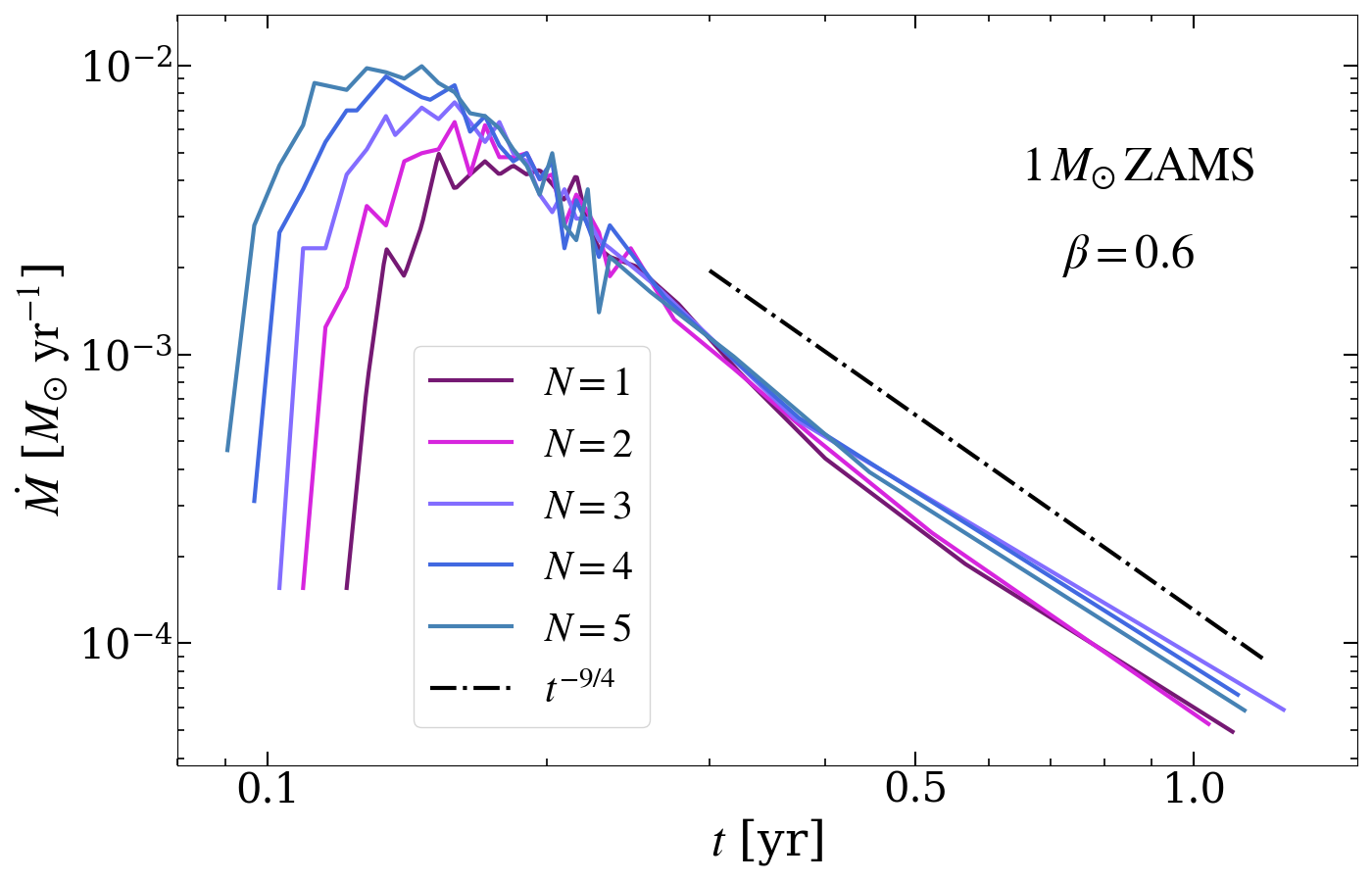}
     \caption{Top: The amount of mass stripped, $\Delta M$, as a function of the number of pericenter passages, $N$, for a $1 M_{\odot}$ ZAMS star with $\beta=0.6$ for each encounter. The absence of a centrally concentrated core-like structure leads to a progressive increase in the amount of mass stripped with the number of pericenter passages. Bottom: Fallback rates for the first five encounters between the $1 M_{\odot}$ ZAMS star and a $10^6 M_{\odot}$ SMBH, at $\beta=0.6$, with a resolution of $10^6$ particles. The peak of the fallback rate increases gradually as the amount of mass stripped increases, and the peak timescale shifts to earlier times.}
     \label{fig:1msunZAMS-beta0p6}
\end{figure}
The density profile of the $1 M_{\odot}$ ZAMS star is not as centrally concentrated as that of a high-mass and evolved star. Consequently, the critical impact parameter $\beta_{\rm c}$ required for the complete disruption of the star is $\sim 1.8$ -- which is approximately a factor of 3 smaller than that required to completely destroy a $3 M_{\odot}$ TAMS star through tidal interactions with a $10^6 M_{\odot}$ SMBH. \cite{coughlin22} estimated the minimum value of the impact parameter $\beta$ at which a star begins to lose any mass to be $\beta_{\rm partial} \simeq 0.6$, independent of the stellar properties. To simulate a grazing encounter between a $1 M_{\odot}$ ZAMS star and a $10^6 M_{\odot}$ SMBH, we chose an impact parameter $\beta=0.6$. The top panel of Figure~\ref{fig:1msunZAMS-beta0p6} shows the amount of mass stripped in these simulations as a function of the number of pericenter passages. For this star, the lack of a centrally concentrated core-envelope structure leads to an increasing amount of mass being stripped at each successive pericenter passage. The bottom panel of Figure~\ref{fig:1msunZAMS-beta0p6} shows the fallback rates for the first five encounters. As seen in the figure, the peak timescale decreases as the number of pericenter passages increases. The magnitude of the peak of the fallback rate gradually increases, which can be attributed to the increase in the amount of mass lost as well as the decrease in the peak timescale. This result shows that young low-mass stars that are on grazing orbits around an SMBH are unlikely to give rise to repeated flares of comparable peak luminosity, as observed for ASASSN-14ko.

\begin{figure}
    \includegraphics[width=0.48\textwidth]{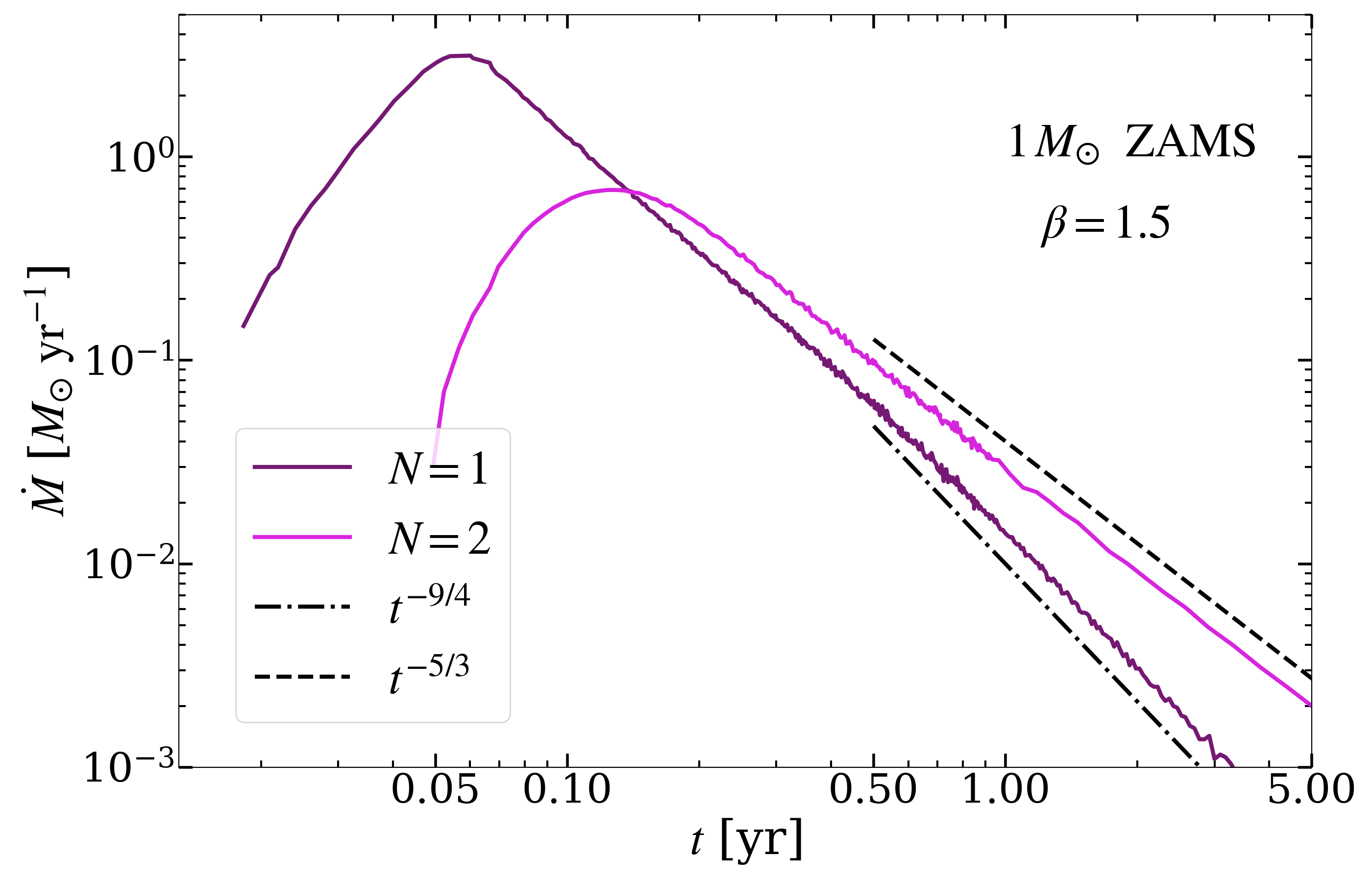}
     \caption{Fallback rates for the disruption of the $1 M_{\odot}$ ZAMS star at $\beta=1.5$. The first encounter leads to a partial disruption, but strips of a significant fraction of the mass of the star. The second encounter results in the complete disruption of the star, giving rise to a fallback rate exhibiting a smaller peak magnitude, that scales as $\propto t^{-5/3}$ at late times. }
    \label{fig:fbr-1msun-beta1p5}
\end{figure}

Finally, to explore the other end of the spectrum of $\beta$ values pertinent to partials, we simulated two successive disruptions of the  $1 M_{\odot}$ ZAMS star at $\beta=1.5$, for which the fallback rates are shown in Figure~\ref{fig:fbr-1msun-beta1p5}. As noted above, the critical impact parameter for the complete disruption of this star is $\beta_{\rm c} \approx 1.8$~\citep{golightly19b,coughlin22}, and thus for $\beta=1.5$, the star loses a significant fraction of its mass during the first pericenter passage. From the figure, we see that the late-time scaling of the fallback rate for the first encounter scales as $\propto t^{-9/4}$, indicating that the disruption is partial. However, the star is completely destroyed on its second pericenter passage, yielding a fallback rate that has a lower peak magnitude relative to the first, and that scales as $\propto t^{-5/3}$ at late-times. The dimmer second peak is consistent with the observed lightcurve for AT2018fyk~\citep{wevers23}. {Since the second encounter results in a complete disruption of the star, if AT2018fyk was generated by a low-mass star on an orbit having a $\beta$ value close to that required for its complete disruption, we would not expect to observe any more flares from this source. However, the detection of a second prompt shutoff implies that this explanation likely fails for AT2018fyk, and suggests a third rebrightening in 2027~\citep{pasham24b}}.

\section{Discussion and Conclusions}
\label{sec:conclusions}

\subsection{Spin-up of the stellar core}
\label{sec:spinup}
In Section~\ref{sec:lagrangian} we used a simple toy model of a ``dumbbell'' (composed of two point masses connected by a rigid rod) gravitationally interacting with a black hole to argue that a star should not be torqued beyond a rotational velocity of $\Omega_{\rm p} = \sqrt{G M_{\bullet}}/r_{\rm p}^{3/2}$. While this outcome and toy model agreed qualitatively with our hydrodynamical simulations, there were important differences. Specifically, instead of being torqued to $\sim \Omega_{\rm p}$ on the first encounter with a pseudo-random oscillation in its rotational velocity thereafter, the star was imparted a small change (which declined with the number of pericenter passages; see Figure \ref{fig:angularmomentum}) in $\Omega$ each time it interacted strongly and tidally with the SMBH. This can be attributed to both an increase in the size of the star, such that the star is more efficiently gravitationally torqued by the SMBH, and the fact that a renewed quadrupole moment is excited on each encounter, the nonlinear interaction of which with the tidal field of the SMBH is ultimately and at least partially responsible for generating the net rotation. Our results for the $3 M_{\odot}$ TAMS star on a $\beta=1$ orbit show that the angular velocity imparted to the surviving stellar core over multiple encounters with an SMBH is a small fraction of its break-up spin, and asymptotes to a value of $0.07 \Omega_{\rm c}$ in the limit of a large number of pericenter passages.

We note that while the numerical fit to the rotational velocity of the core implies an asymptotic rotation rate of $\sim 1.4 \Omega_{\rm p}$, we expect the value to asymptote to only $\sim \Omega_{\rm p}$. As discussed in Section \ref{sec:lagrangian}, the reason for this is that once the rotation rate of the star is $\sim \Omega_{\rm p}$, the SMBH appears stationary in the corotating frame of the star near pericenter (which is the only location at which the tidal field is sufficiently strong to torque the star). There is thus no misalignment between the tidal bulge and the vector joining the COM of the star and the SMBH in this frame, meaning that the star will not be torqued beyond this value. 

We also emphasize that while the star should be asymptotically torqued to $\Omega_{\rm p}$ if it is initially not rotating, the outcome is qualitatively different if the star is initially rotating faster than $2\Omega_{\rm p}$ in a prograde sense or faster than $\Omega_{\rm p}$ in a retrograde sense. In these cases, the SMBH makes more than one complete rotation in the corotating frame of the star near pericenter, meaning that the net tidal torque is much weaker and the star and its rotation is relatively unperturbed. This limit of super-Keplerian rotation is applicable to rpTDEs if the Hills mechanism is responsible for placing the star on its tightly bound orbit, which is a viable mechanism for producing such short-period orbits if there were not many prior encounters (\citealt{cufari22b}; as was mentioned in Section \ref{sec:intro}, this condition is necessary for systems such as ASASSN-14ko that lose $\sim 1\%$ of the total stellar mass per encounter). However, even {with} the Hills mechanism, extremely tight binary separations of $\sim few \times R_{\odot}$ are required to produce $\sim 100$-day orbits about SMBHs with masses $\gtrsim 10^{6-7}M_{\odot}$, such that the binding energy of the binary is comparable to that of the partially disrupted star. We would then expect the star to be tidally locked at a rotation rate that can, depending on the pericenter distance of the binary and its tidal disruption radius, exceed $\Omega_{\rm p}$, in which case the rotation of the star would be largely unaffected. Because the binary can be randomly oriented with respect to the orbital plane of the binary center of mass, the result would be a rapidly rotating star at an inclined angle about the SMBH.

Finally, the rotation imparted to the star is a nonlinear effect: the tidal bulge provides the quadrupole moment, which subsequently experiences the torque applied by the tidal field. Additionally, \cite{kochanek92} argued that the tidal interaction excites modes that have non-zero vorticity, and the damping of these modes establishes a bulk rotation within the star, such that the angular momentum in the oscillatory modes and that in the rotation of the star cancel out the net vorticity. We discuss the development of solid-body rotation within the star in Appendix~\ref{sec:rotation} and show that it is insensitive to numerical resolution, i.e., it is a physical consequence of the tidal interaction with the SMBH. Because it is a nonlinear effect, we would expect the net rotation per encounter to decline extremely strongly with distance from the SMBH, similar to the way in which the energy imparted via tides falls off with $\beta$ very rapidly \citep{press77}. Thus, the statement that the star is spun up to $\sim \Omega_{\rm p}$ is likely only valid for $0.5 \lesssim \beta \lesssim \beta_{\rm c}$, where $\beta_{\rm c}$ is the value of $\beta$ at which the star is completely destroyed. 

\subsection{Effect of Stellar Structure on Survivability}
Our hydrodynamical simulations show that the survivability of a star undergoing repeated tidal interactions with an SMBH depends on its structural properties. 
High-mass stars that are in their late evolutionary stages develop a core-envelope structure that enhances their probability of survival relative to low-mass, less centrally concentrated stars. This is in agreement with the results of ~\cite{liu23}, where the authors concluded, based on an adiabatic mass-loss model, that a centrally concentrated high-mass star having a diffuse outer envelope can generate ASASSN-14ko like flares as it undergoes periodic mass-loss events on a grazing orbit around an SMBH. We showed that the $3 M_{\odot}$ TAMS star and the $1.3 M_{\odot}$ TAMS star (for which the critical impact parameter required for their complete disruption is $\beta_{\rm c} \simeq 6$) on a $\beta=1$ orbit can survive many encounters, losing a small amount of mass on each pericenter passage, and generating luminous flares for a long time. On the other hand, for the $1 M_{\odot}$ ZAMS star, the density profile is not as centrally concentrated, and the amount of mass stripped per encounter is a monotonically increasing function of the number of pericenter passages, which can be detrimental to the survival of the star. \cite{liu24a} studied the hydrodynamical evolution of the disruption of a sun-like star with $\beta=0.5, 0.6 \, \rm and \, 1.0$, and found that the star loses an increasing amount of mass with each pericenter passage, and is destroyed in $\lesssim 10$ encounters for $\beta=0.5$, to $\lesssim 3$ encounters for $\beta=1.0$. While the $1 M_{\odot}$ star studied in their simulations is more evolved than the ZAMS star, our results for the $1 M_\odot$ ZAMS star are qualitatively in agreement with their result, that a sun-like star loses an increasing amount of mass with each pericenter passage, and is destroyed within a few orbits. The same authors also note that the presence of a centrally concentrated core and a tenuous outer envelope enables a more evolved star to survive a greater number of tidal encounters without being completely destroyed, which is consistent with our results for the $3 M_\odot$ TAMS star.

\cite{antonini11} used $N-$body integrations and SPH simulations to study the repeated tidal stripping of main-sequence of stars as one of the possible outcomes of the tidal break-up of binary star systems by SMBHs. They used $N-$body integrations to evolve the binaries for $\gtrsim 100$ orbits around the SMBH, and found that in a Hills-capture scenario, where one of the stars is ejected as a hyper-velocity star, the bound companion can undergo periodic mass-loss episodes for hundreds of orbits before being completely destroyed. Their SPH simulations, which have a maximum resolution of $\sim 4 \times 10^4$ particles, do not explore the parameter space of stellar and orbital parameters that we considered here, but our results qualitatively agree with their conclusion -- that main-sequence stars on highly eccentric bound orbits around SMBHs can undergo repeated tidal disruptions, giving rise to periodic flares of varying intensity. We demonstrated that the strength, duration and periodicity of these flares are largely determined by the structural properties of the star and the distance of closest approach between the star and the SMBH. 

\cite{zalamea10} studied the feasibility of detecting electromagnetic and gravitational-wave signals arising from an extended period of slow mass-loss in white dwarfs on inspiralling orbits around a $10^5 M_{\odot}$ SMBH. While the region of parameter space explored in their work does not directly correlate to ours, they found that white dwarfs on highly eccentric orbits around an SMBH, undergoing slow periodic mass-loss episodes, can last for thousands of orbits, emitting gravitational-waves detectable in the Laser Interferometer Space Antenna (LISA) band and accretion flares with luminosities close to the Eddington limit of the SMBH.~\cite{macleod13} used a combination of hydrodynamical simulations and a semi-analytic model to study the tidal stripping of a giant star with mass $M_{\star}=1.4 M_{\odot}$ and $R_{\star}=50 R_{\odot}$, on an eccentric orbit around a $10^7 M_{\odot}$ SMBH. The mass-transfer process is initiated when the star evolves up the red giant branch. They showed that the red-giant star can survive for hundreds of orbits around the SMBH, giving rise to low-intensity flares that recur on the orbital timescale of the star, which is $\gtrsim 10^3$ years.

\subsection{Evolution of the Orbital Period and the Peak Fallback Timescale}
\label{sec:evolution}
The secular evolution of the orbital period of ASASSN-14ko is not well understood, and various mechanisms have been suggested for reproducing the observed period derivative of $\dot{P} = -0.0026$. \citet{payne21} and \cite{cufari22b} concluded that gravitational wave emission underestimates the period derivative by several orders of magnitude. \cite{linial24} analyzed various physical processes that could be responsible for producing the observed period change in ASASSN-14ko, and suggested that the hydrodynamical drag experienced by the star as it interacts with the accretion disc of the SMBH provides the most likely explanation for the observed change in period. However, they noted that the minimum disc mass required for this likely exceeds the mass of the star, thereby necessitating the existence of an AGN-like disc to explain the observed evolution of the period. 

Our analytical estimates from Section~\ref{sec:lagrangian} show that for solar-like stars being repeatedly tidally disrupted by an SMBH, the angular velocity imparted to the stellar core can change the period by a small fraction of its original value. For the specific case of ASASSN-14ko, having a period of $114\pm1$ days, we estimated the fractional change in the period to be $\Delta T / T \simeq -0.003$, which is in excellent agreement with observations \citep{payne23}. We note, however, that to continue to reduce the orbit by the same fractional amount, this energy must be lost from the star and efficiently, i.e., on a per-orbit basis. Since the imparted angular momentum to the star results in an angular velocity that is approximately uniform, it is unlikely that this energy will be dissipated efficiently, meaning that -- if the imparted rotation does represent the sink of orbital energy that leads to the period decay -- one would expect a pronounced $\ddot{P}$ as the star is not efficiently torqued above $\Omega_{\rm p}$.

In addition to bulk rotation, one expects a comparable amount of energy produced in oscillatory modes that is ultimately dissipated as heat, either viscously or through three-mode (and higher) nonlinear couplings (e.g., \citealt{mcmillan87, kochanek92, kumar96, weinberg12}). As we discuss in Appendix~\ref{sec:tidal heating}, we expect most of the thermal energy to be concentrated in the outer layers of the star where the amplitude of the tidal acceleration is largest, and these layers are removed on subsequent encounters with the SMBH. For ASASSN-14ko (i.e., with an orbital period of 114 days and a SMBH mass of $10^7M_{\odot}$), the reduction in the orbital period would result in the deposition of $\sim 1.5\times 10^{46}$ erg in $0.01 M_{\odot}$ (i.e., comparable to the amount of mass stripped from the star in our simulations), which would require a luminosity of $L\approx 3.5 L_{\rm Edd}$ if this energy were to be lost radiatively ($L_{\rm Edd}$ is the Eddington luminosity of a $3M_{\odot}$ star, assuming $\kappa_{\rm es} = 0.34$ cm$^2$ g$^{-1}$ for the opacity; see the discussion in Appendix~\ref{sec:tidal heating} for more details). The amount of energy dissipated via tides and required to reproduce the observed $\dot{P}$ therefore could not be efficiently exhausted radiatively (see also \citealt{kumar98}), but the tidal stripping of the outer layers of the star serves an alternative outlet through which the energy can be lost mechanically and without destroying the high-density (and thermally insulated; again, see Appendix \ref{sec:tidal heating}) core, thus allowing periodic accretion events to continue. While our resolution tests and the implementation of alternative thermodynamic treatments, discussed in detail in Appendix \ref{sec:tidal heating}, provide evidence to suggest that this is the means by which the star can continue to survive many tidal encounters with the SMBH, additional investigations -- particularly in the context of linear tidal theory -- are required to solidify this connection.

\cite{liu24} argued that the rapid evolution of the recurrence time of the flares observed from eRASSt-J045650, which has now flared $\sim 5$ times, can be interpreted as a change in the orbital period of a $1 M_{\odot}$ star undergoing repeated partial disruptions on an orbit around a $10^5 M_{\odot}$ SMBH. However, they estimated that the amount of mass that the star must lose on a single pericenter passage is $\sim 0.8-0.9 M_{\odot}$. Given that this is a substantial fraction of the original mass of the star, it seems unlikely that the remnant of this interaction would survive subsequent pericenter passages without undergoing a drastic and unphysical change in its pericenter distance. Additionally, as argued in \citet{cufari22b} and as shown here in Section \ref{sec:lagrangian}, reducing the period by the amount claimed in \citet{liu24} would necessitate imparting more energy into the star than its own binding energy. 

While the change in orbital period induced by the spin-up of the stellar core can only be an extremely small fraction of the original period, the peak timescale for the fallback rate of tidally stripped stellar debris can exhibit greater variation due to the imparted rotation. We found from the simulation of the partial disruption of a $3 M_{\odot}$ TAMS star by a $10^6 M_{\odot}$ SMBH that the peak fallback time scales with the imparted angular velocity $\Omega$ approximately as $\left( 1+ \sqrt{2}\Omega/\sqrt{G M_{\bullet}/r_{\rm p}^3}\right)^{-0.8}$. From Figure~\ref{fig:tmb_vs_omega} we see that if an rpTDE is observed during the first few pericenter passages of the star, the peak timescale could show significant variation between successive flares, but as the angular momentum imparted to the stellar core converges to a nearly constant value over multiple encounters, the variation in the peak timescale would be less prominent for later pericenter passages.

\subsection{Connecting ASASSN-14ko's outbursts to orbital and stellar properties}

The fallback rates obtained from our hydrodynamical simulations for high-mass stars qualitatively reproduce the nature of the observed flares from ASASSN-14ko, for which the peak magnitude of the flares shows very little variation over the ten-year period for which this event has now been observed~\citep{payne21,payne22,payne23}. However, the eccentricity of the bound orbit will have imprints on the morphology of the fallback rates that are not captured in our simulations, for which we assumed the COM of the star to be on a parabolic orbit. The hydrodynamical evolution of stellar TDEs on eccentric orbits and the fallback rates of the bound stellar debris have been studied in, e.g., \cite{hayasaki13,park20,cufari22a}. In particular, features such as the peak timescale and the late-time scaling of the fallback rates will differ for an $e \simeq 0.98$ orbit, relative to its parabolic counterpart. Also, since we did not self-consistently evolve the star on its elliptic orbit, we cannot directly constrain the orbital period of the star or the time between successive peaks from our simulations. Thus, we do not attempt to translate the fallback rates for successive encounters along the time axis to depict their chronology of occurrence, but instead overlay them on top of one-another, demonstrating the relative intensity of the peak magnitudes between successive pericenter passages. For any given fallback rate, the time axis represents the lapse of time since the corresponding pericenter passage of the star. The peak timescales for the high mass stars that we simulated at $\beta=1$ ranges from $\sim 40-80$ days. The SMBH mass used in our simulations was $10^6 M_{\odot}$, whereas the inferred SMBH mass for the host galaxy of ASASSN-14ko is estimated to be $\sim 10^7 M_{\odot}$~\citep{payne21}. The peak timescale for TDEs scales with the mass of the SMBH as $\propto M_{\bullet}^{1/2}$~\citep{lacy82}, and so the expected peak timescale for ASASSN-14ko's flares should be longer, based on its SMBH mass estimate. Since the observed UV lightcurve rises to its peak on a timescale of $\sim 2 $ days~\citep{payne22}, much of the early time accretion rate is presumably not observed.

For observed events, the recurrence timescale is generally defined as the time between successive peaks in the lightcurve. This timescale does not directly correlate with the orbital or stellar parameters. Our ability to constrain these parameters depends on the presence or absence of certain features in the observed lightcurve. Specifically, there will be some time-lapse between the sequence of events that constitute the process of disruption and the ensuing accretion onto the SMBH, namely, the star reaching pericenter, the initiation of accretion, the rising phase of the lightcurve, and its subsequent decay. If the timescale on which the fallback rate rises and decays is comparable to or longer than the time between successive peaks, then the returning of the stellar core to the pericenter would be associated with a prompt shutoff of the fallback rate~\citep{liu23,wevers23}, allowing us to constrain the orbital period of the star. Such a prompt shutoff was observed in the X-ray lightcurves of transients such as AT2018fyk and eRASSt-J045650~\citep{wevers23,liu24}. On the other hand if, as is the case for ASASSN-14ko, the lightcurve decays on a timescale that is much shorter than the time between successive peaks, then the orbital period of the star can not be constrained from the observed lightcurves of the events. In addition to orbital parameters, rpTDE lightcurves can also allow us to constrain the properties of the disrupted star. As we showed here, ASASSN-14ko's outbursts are consistent with the partial disruption of a high-mass star with a centrally concentrated core and a diffuse outer envelope, on a grazing orbit around a SMBH, losing $\sim 0.01 M_{\odot}$ on each pericenter passages, and surviving multiple encounters to give rise to the observed outbursts.

\section{Summary}
\label{sec:summary}
We used analytical estimates based on a simplified toy model, and hydrodynamical simulations of the repeated partial tidal disruption of stars by an SMBH, to analyze the feasibility of the rpTDE model for generating repeating nuclear transients. The key findings of our work are:

\begin{enumerate}
\item{A high-mass star in its late evolutionary stages can undergo repeated mass-transfer events on a bound orbit around an SMBH, losing a small fraction of its outer envelope on successive pericenter passages, giving rise to ASASSN-14ko-like flares over a sustained period.}

\item{The relative brightness between successive peaks of observed lightcurves is determined, to a large extent, by the type of star and the impact parameter characterizing the orbit. This can be used to constrain the properties of the disrupted star from the observed lightcurves of rpTDEs.} 

\item{Using the energy-period relation for the Keplerian orbit, we showed that the rotational velocity imparted to the surviving stellar core can lead to changes in its period that are comparable to the observed period derivative of $\dot{P}=-0.0016$ for ASASSN-14ko, provided the star is spun up to a rotational velocity comparable to $\sim \Omega_{\rm p}$ in one orbit. Our simulations show (see Figure \ref{fig:angularmomentum}), however, that it takes several pericenter passages to spin up the star by this amount, and thus the rotational energy of the core does not act as an efficient sink for the dissipated orbital energy (at least as concerns the observed $\dot{P}$ for ASASSN-14ko). Moreover, the per-orbit increment in $\Omega$ declines with time and the star can not be spun up beyond $\sim \Omega_{\rm p}$, implying that there would be a significant $\ddot{P}$ if this were the primary source of dissipation. Instead, it is likely (and consistent with the results in Appendix \ref{sec:tidal heating}) that tidal dissipation deposits thermal energy (at the expense of the binding energy of the orbit) in the outer layers of the star that are tidally stripped on subsequent encounters, i.e., the energy is lost mechanically instead of radiatively, while the core is relatively unmodified and survives many more encounters.}

\item{The timescale on which the fallback rate rises and peaks decreases as a power-law in the imparted angular velocity ($t_{\rm peak} \propto \left( 1+ \sqrt{2} \Omega \slash \Omega_{\rm p} \right)^{-0.8}$). It shows greater variation for earlier orbits, and eventually converges to an almost constant value. Thus, if an rpTDE is observed during the first few pericenter passages of the star on its orbit around an SMBH, the timescale on which the lightcurve rises and peaks should show a significant change between successive flares.} 

\item{A non-rotating star is spun up to $\sim \Omega_{\rm p}$ upon interacting many times with the SMBH. If the star is initially rotating faster than $\sim 2\Omega_{\rm p}$ in a prograde sense, or faster than $\sim \Omega_{\rm p}$ in a retrograde sense, which could arise if the star is captured by the black hole through a three-body exchange (i.e., the Hills mechanism), the same conclusion does not hold and the stellar rotation will be relatively unperturbed.}

\end{enumerate}

\section*{}
We thank Itai Linial and Eliot Quataert for useful discussions, and the anonymous referee for useful comments and suggestions that improved the manuscript. A.B.~and E.R.C.~acknowledge support from NASA through the \emph{Neil Gehrels Swift} Guest Investigator Program, proposal number 1922148. E.R.C.~acknowledges support from the National Science Foundation through grant AST-2006684, and from NASA through the Astrophysics Theory Program, grant 80NSSC24K0897. C.J.N. acknowledges support from the Science and Technology Facilities Council (grant No. ST/Y000544/1) and from the Leverhulme Trust (grant No. RPG-2021-380). This research was supported in part by grant NSF PHY-2309135 to the Kavli Institute for Theoretical Physics (KITP).

\appendix
\section{Effects of tidal heating}
\label{sec:tidal heating}
\begin{figure}[h]
    \includegraphics[height=5.5cm,width=0.48\textwidth]{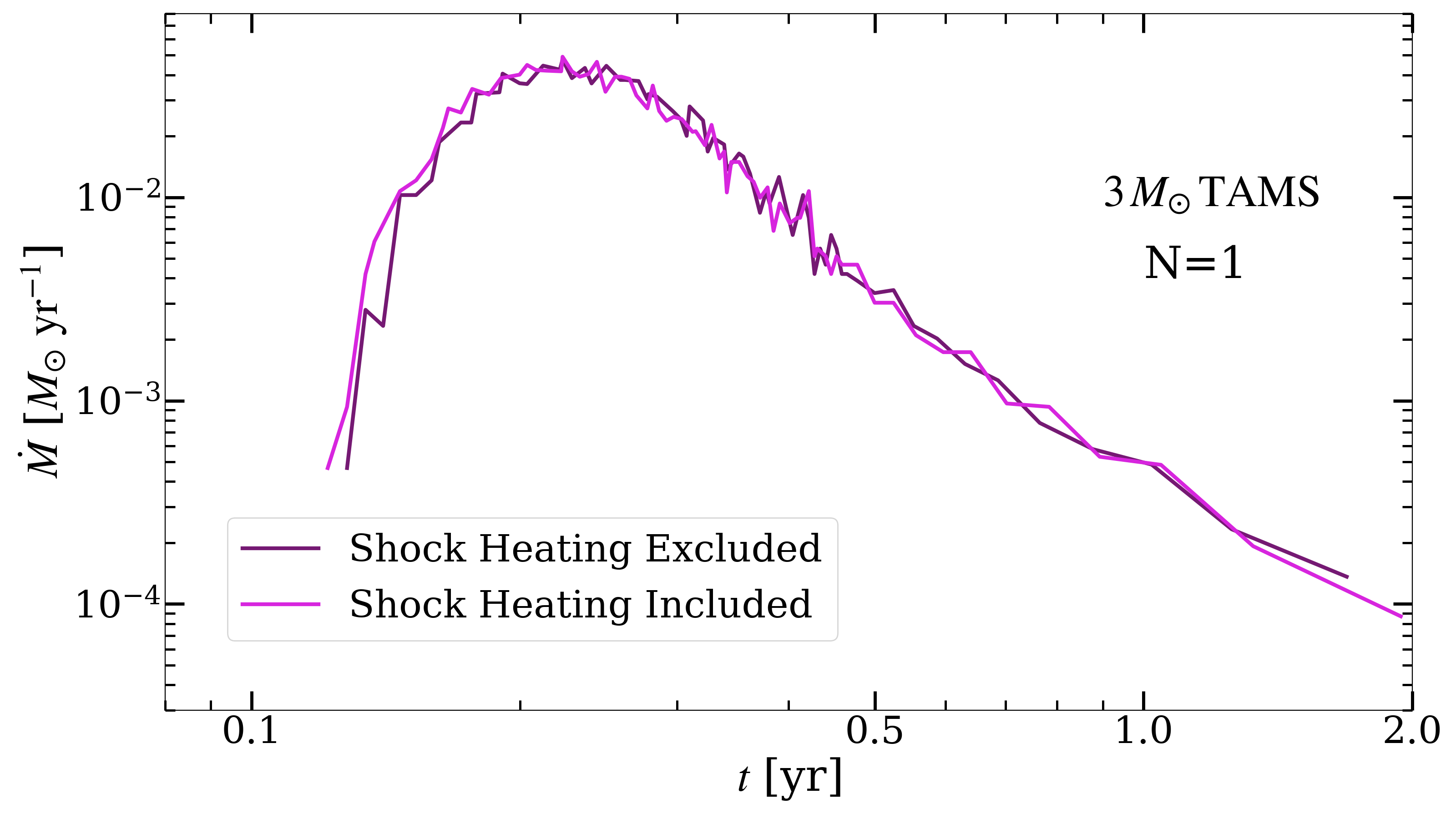}
    \includegraphics[height=5.5cm,width=0.48\textwidth]{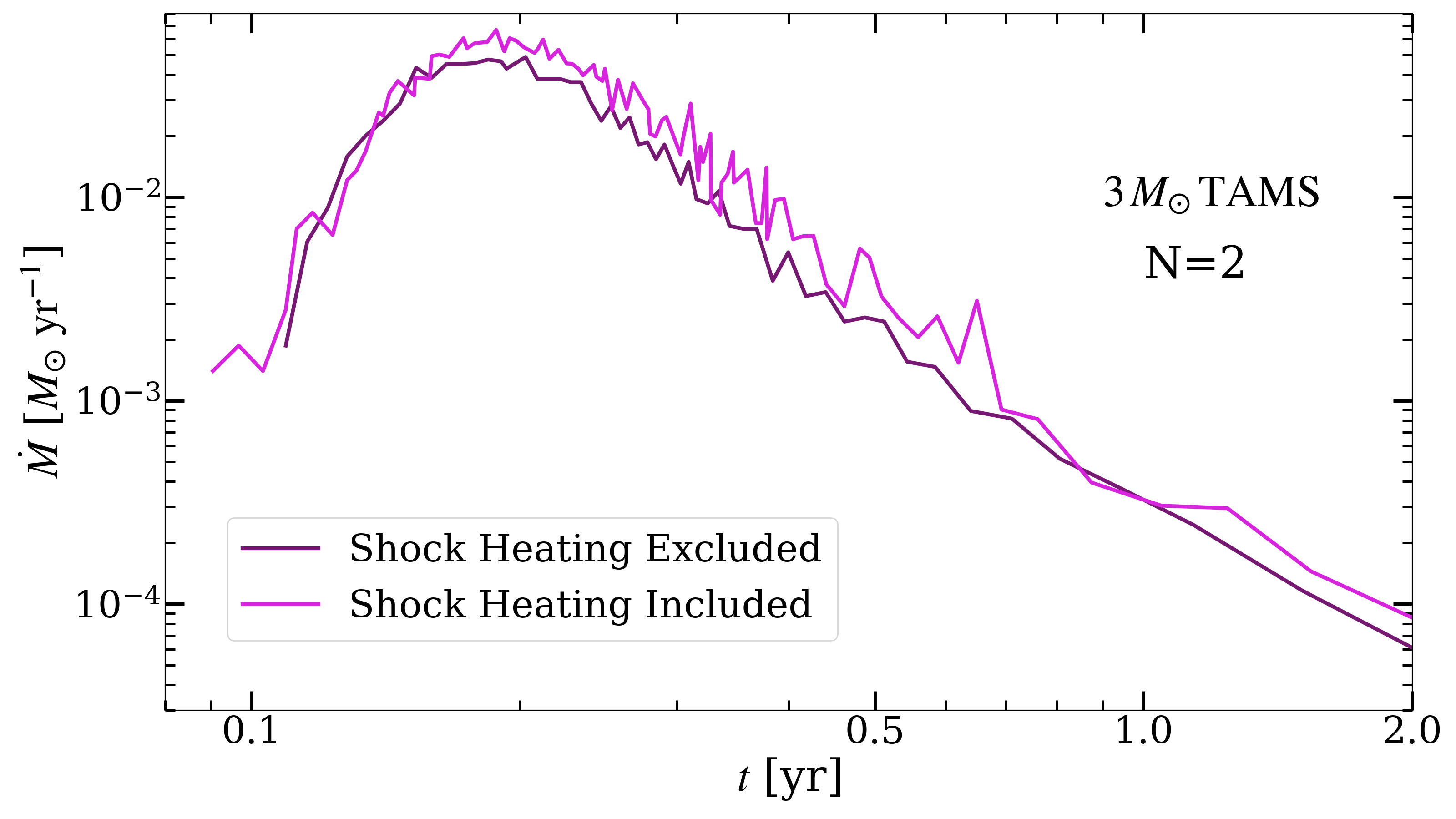}
    \includegraphics[height=5.5cm,width=0.48\textwidth]{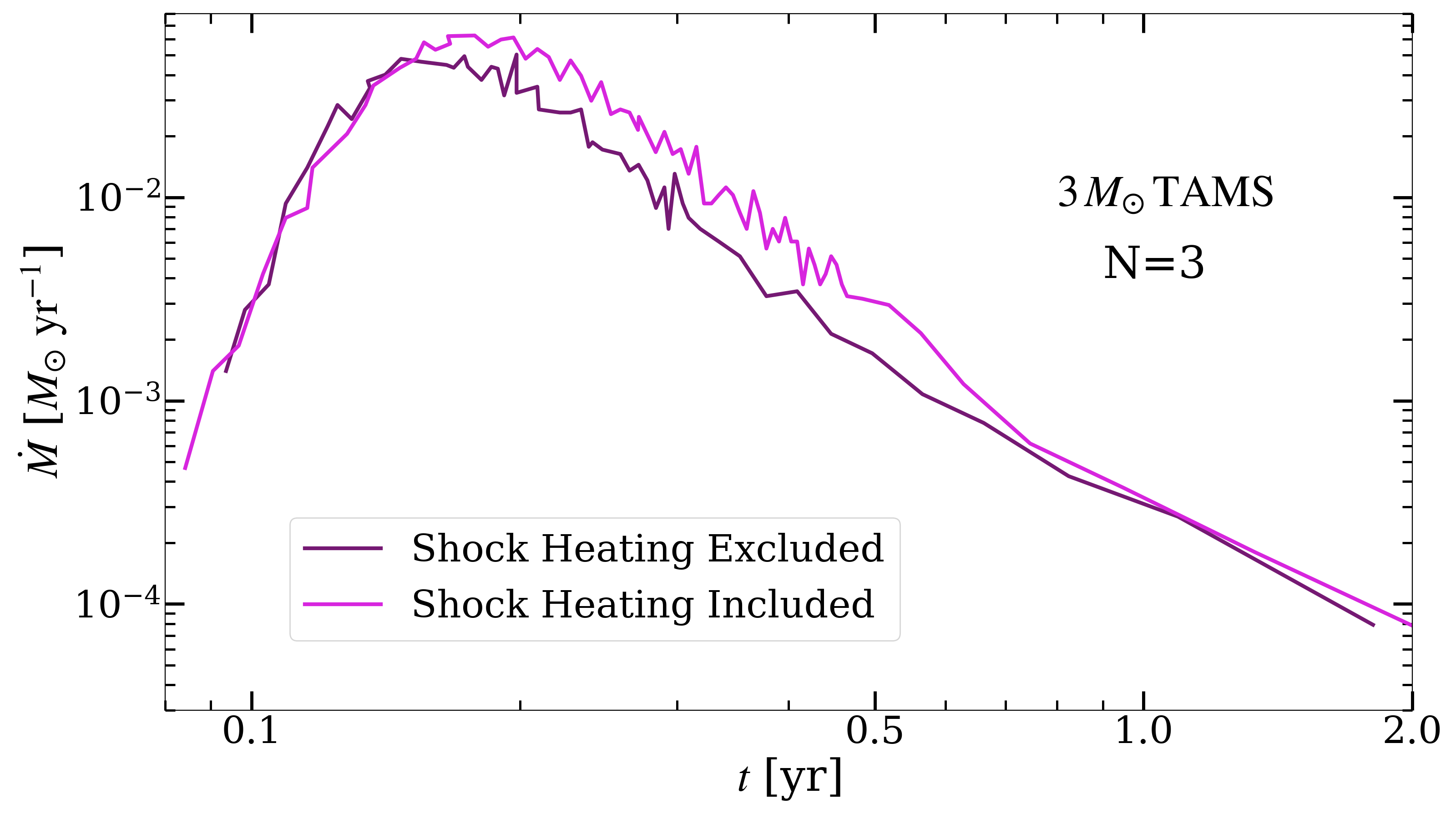} 
    \includegraphics[height=5.5cm,width=0.48\textwidth]{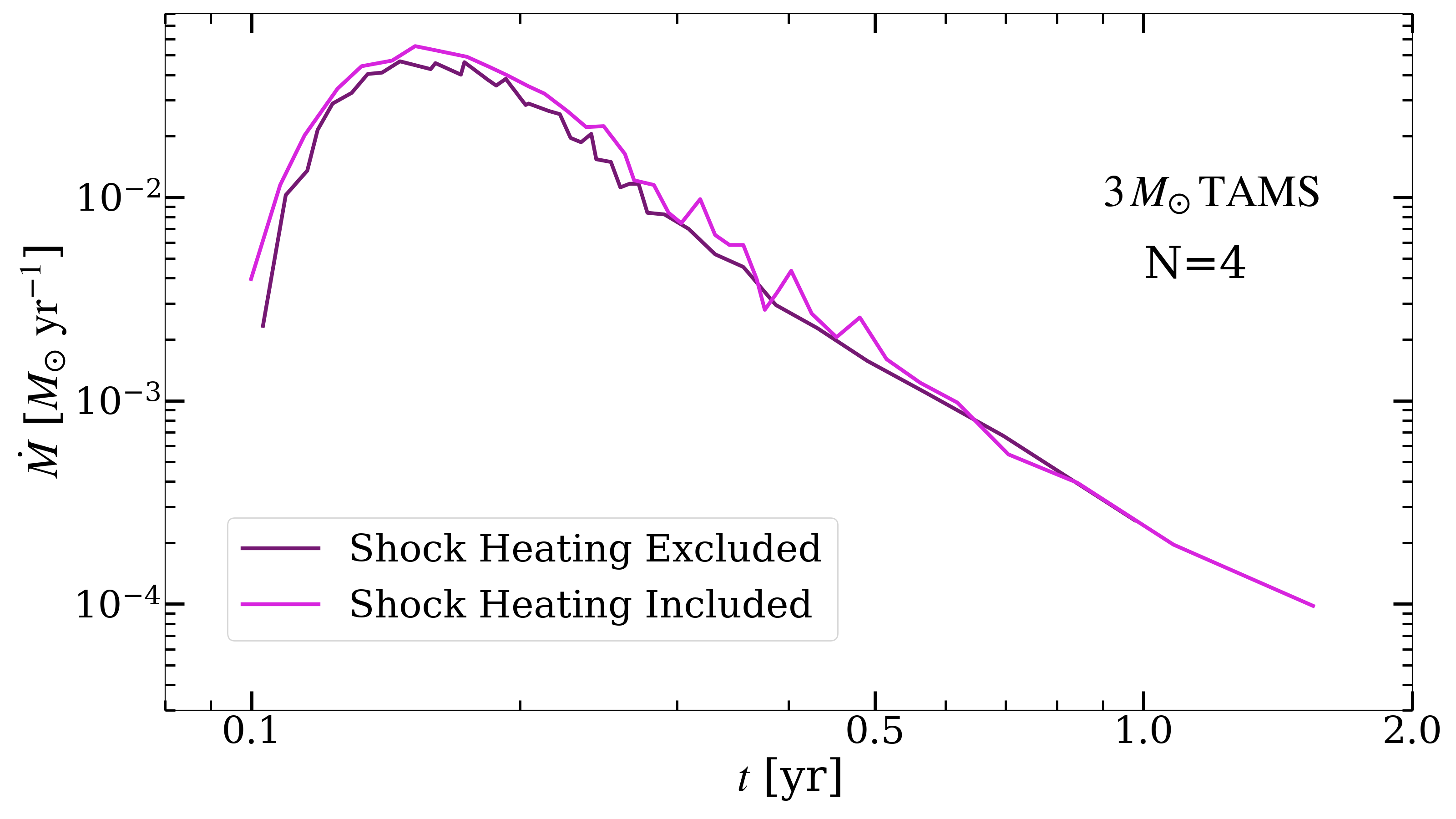}
    \includegraphics[height=5.5cm,width=0.48\textwidth]{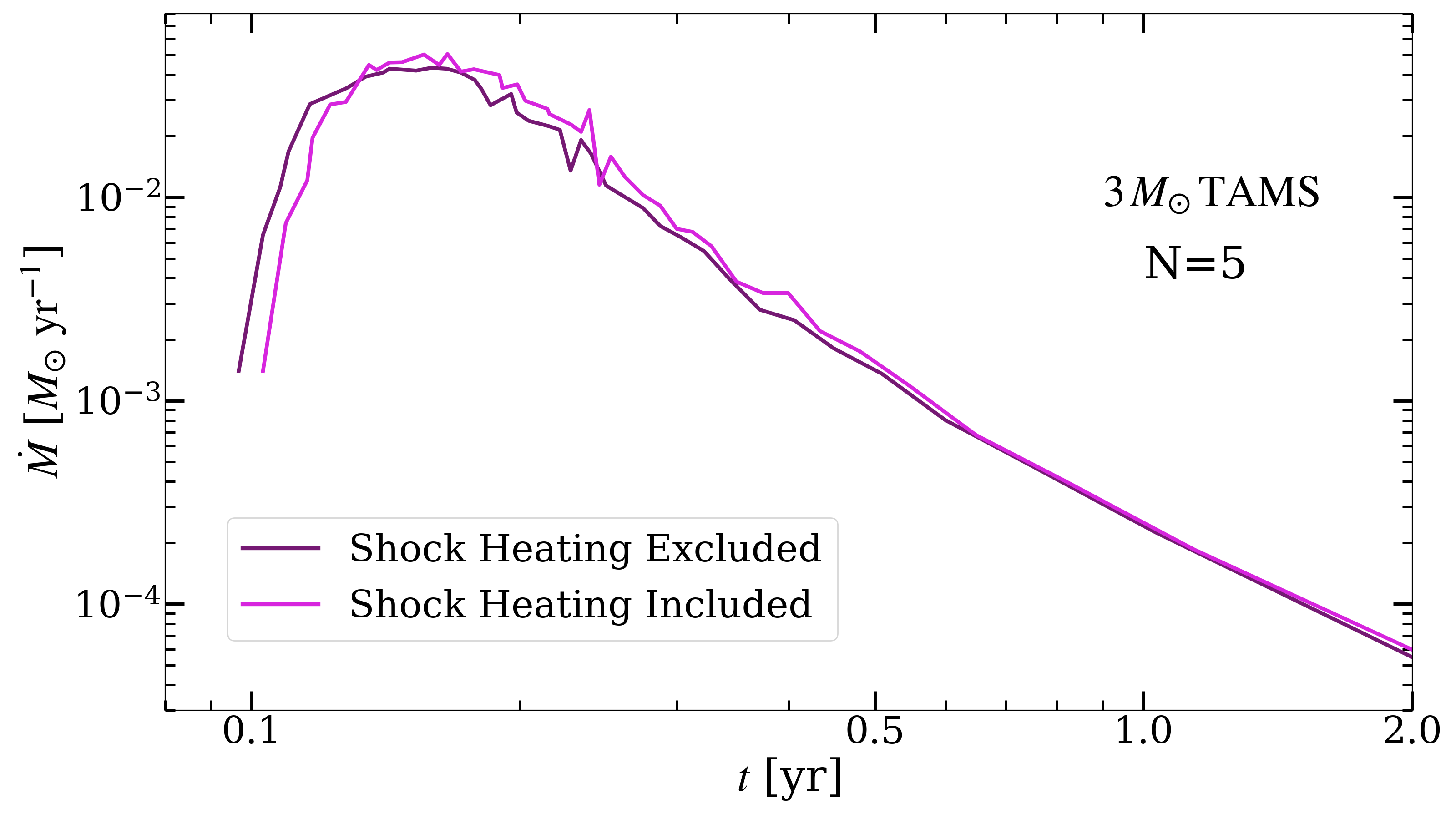}\hspace{6mm}
    \includegraphics[height=5.5cm,width=0.48\textwidth]{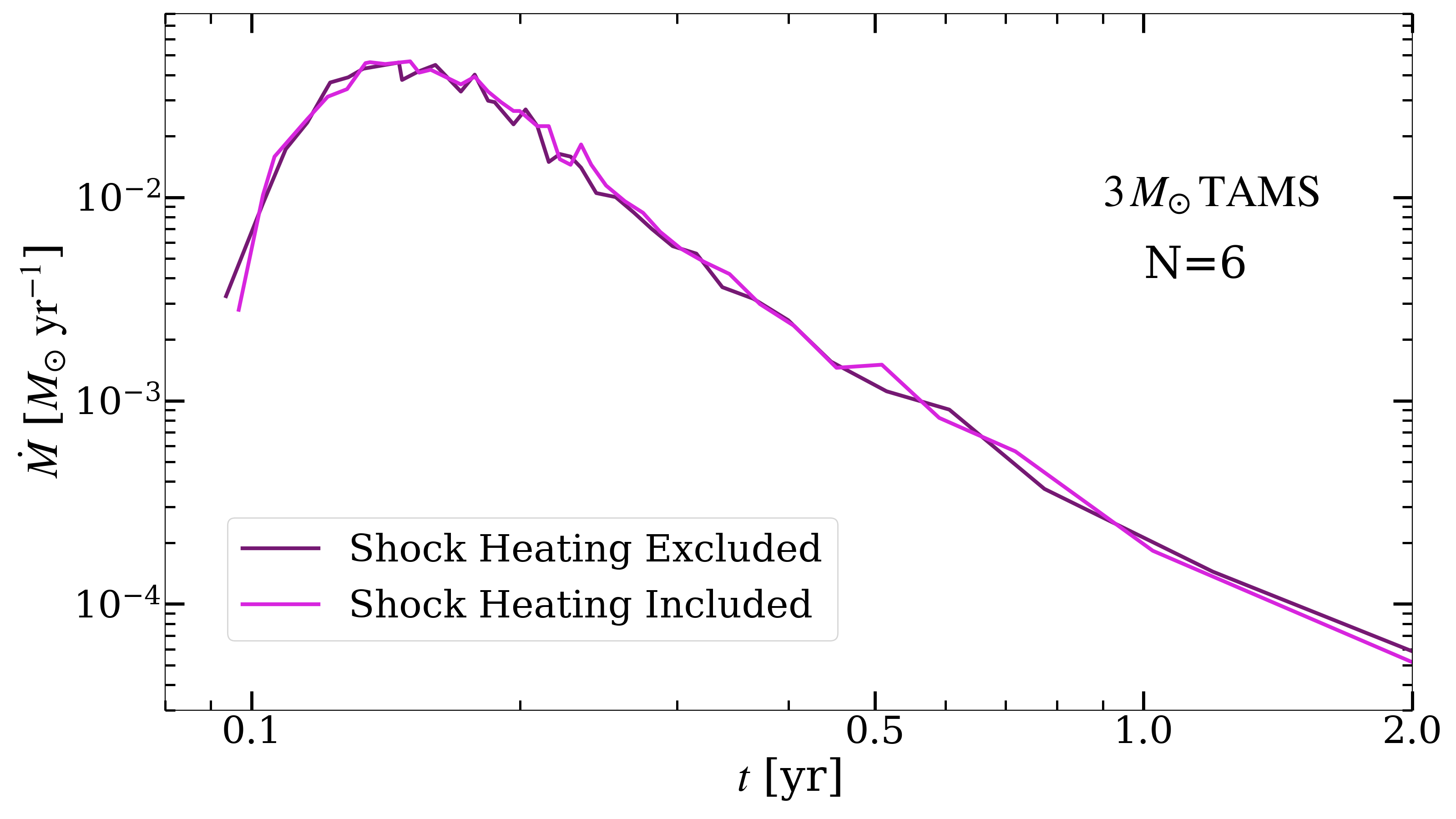}
    
     \caption{Fallback rates onto a $10^6 M_{\odot}$ SMBH from the partial disruption of the $3 M_{\odot}$ TAMS star on its first six pericenter passages, with (pink) and without shock heating (purple). With the inclusion of shock heating, the outer layers of the star puff-up, leading to an increased amount of mass loss as compared to the non shock heating case. The fallback rates for the second and third encounters show the maximum deviation from the non shock heating case. Once the outer layers of the star are removed, subsequent pericenter passages lead to a comparable amount of mass being lost from the surviving core, and almost indistinguishable fallback rates with and without heating. }
     \label{fig:3msun-shock-heating}
\end{figure}
To assess the impact of tidal heating on the survivability of the star, we performed simulations for the $3 M_{\odot}$ TAMS star that are identical to those presented in Section \ref{sec:hydro3M}, except that the thermal energy that is generated as a byproduct of viscous dissipation (i.e., due to the numerical implementation of an artificial viscosity; see \citealt{price18} for details in the context of {\sc phantom} specifically) is retained within the fluid. In contrast, the simulations presented in Section \ref{sec:hydro3M} assume that this heat is lost from the system; which of these scenarios is closer to reality depends on where the energy is deposited in the star, the rate at which the kinetic energy imparted via tides is viscously and radiatively (or nonlinearly; e.g., \citealt{mcmillan87, kochanek92, kumar96, weinberg12}) damped, and the photon mean free path, but in general we expect the numerical heating to be artificially large in the outer layers of the star where the resolution is lowest (see \citealt{norman21, coughlin22b} specifically in the context of TDEs). 

The fallback rates for the first six encounters for the 1M particle simulations, with and without ``shock heating'' (the term we adopt for numerical heating, which is due to the increased importance of viscosity in enforcing the smoothness of the flow, but is not necessarily due to the presence of a shock), are shown Figure~\ref{fig:3msun-shock-heating}. While the fallback rates for $N = 1$ are almost indistinguishable, the retention of heat within the star causes the outer layers of the star to inflate, and mass is lost more easily from these layers relative to the case where the heat is dissipated. This causes an increased amount of mass loss in the second and third pericenter passages, and the fallback rates for these two encounters lie systematically above the corresponding fallback rates from the simulations in which shock heating is ignored. Table~\ref{tab:table1} shows the amount of mass lost by the star and that is accreted onto the SMBH for the first six pericenter passages. Once the tidally heated outermost layers of the star are stripped, shock heating is no longer significant and the fallback rates are again indistinguishable.

\begin{figure}[h]
    \centering
    \includegraphics[height=8.5cm,width=0.8\textwidth]{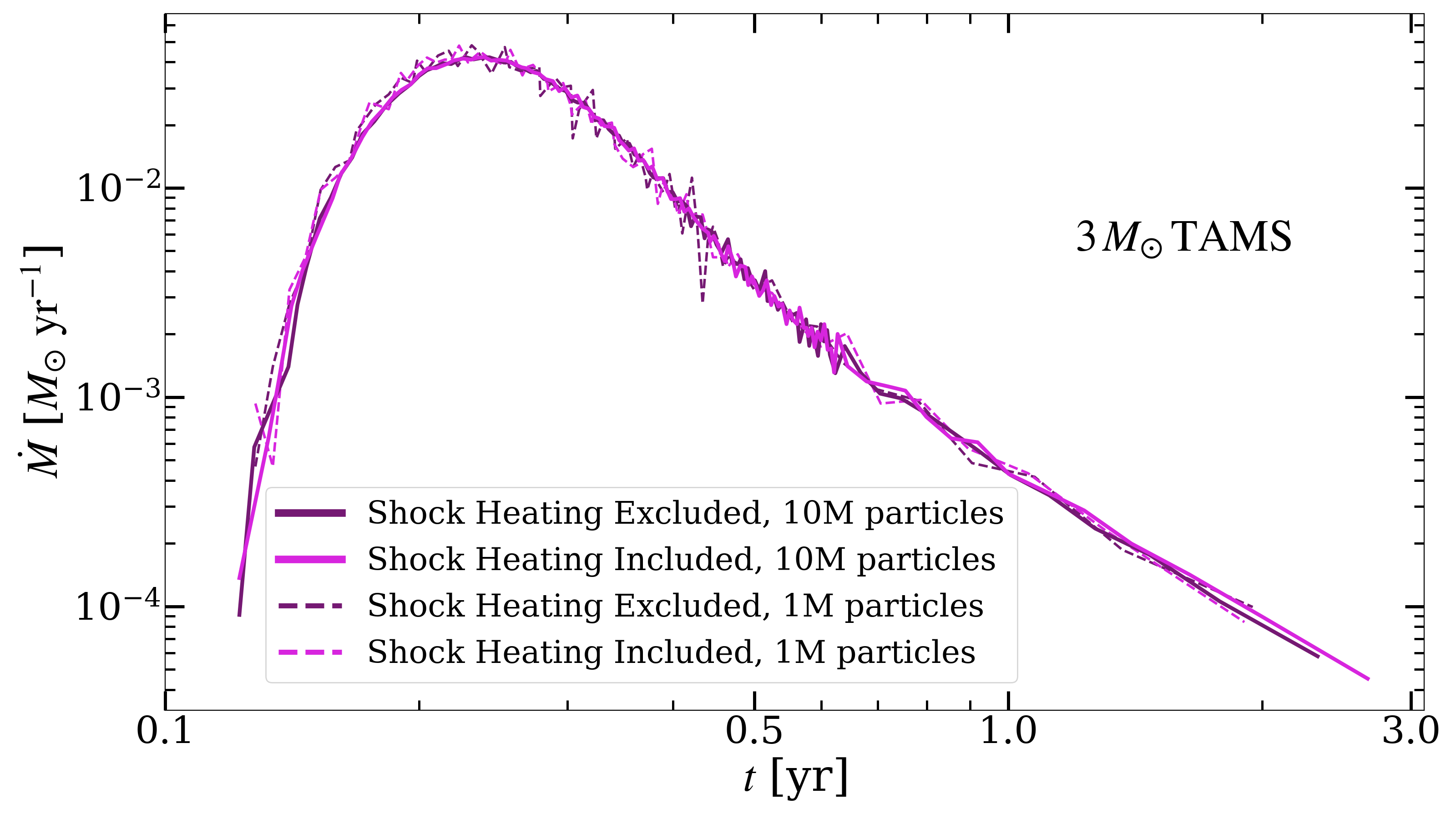}
    
     \caption{The fallback rates for the first pericenter passage of the $3 M_{\odot}$ TAMS star, with and without shock heating. The solid curves are for a higher resolution of 10M particles, whereas the dashed curves are for a resolution of 1M particles.}
     \label{fig:3msun-shock-heating2}
\end{figure}

\begin{figure}[h]
    \advance\leftskip-2cm
    \includegraphics[height=7.5cm,width=0.455\textwidth]{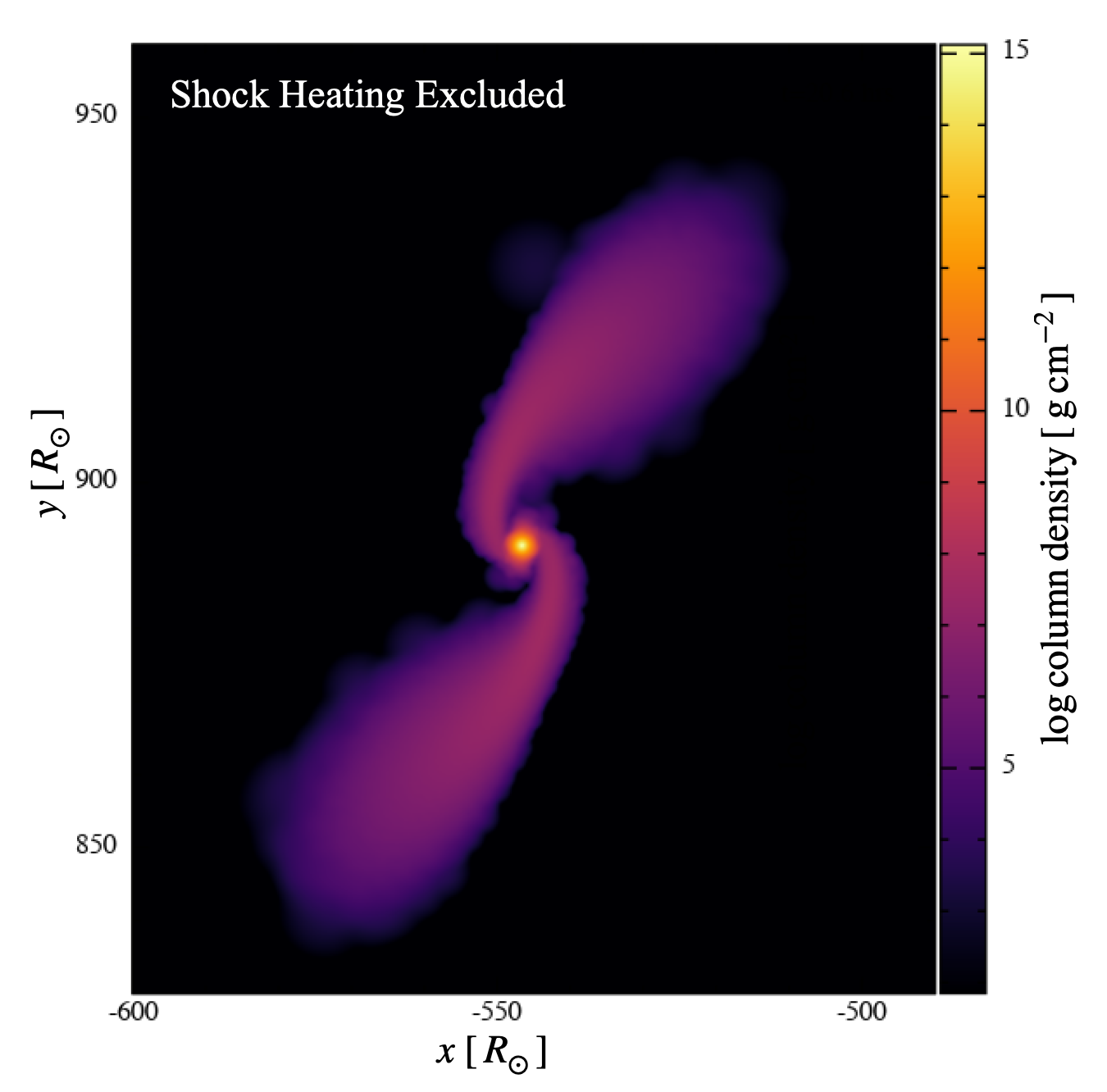} \hspace{0.1mm}
    \includegraphics[height=7.5cm,width=0.455\textwidth]{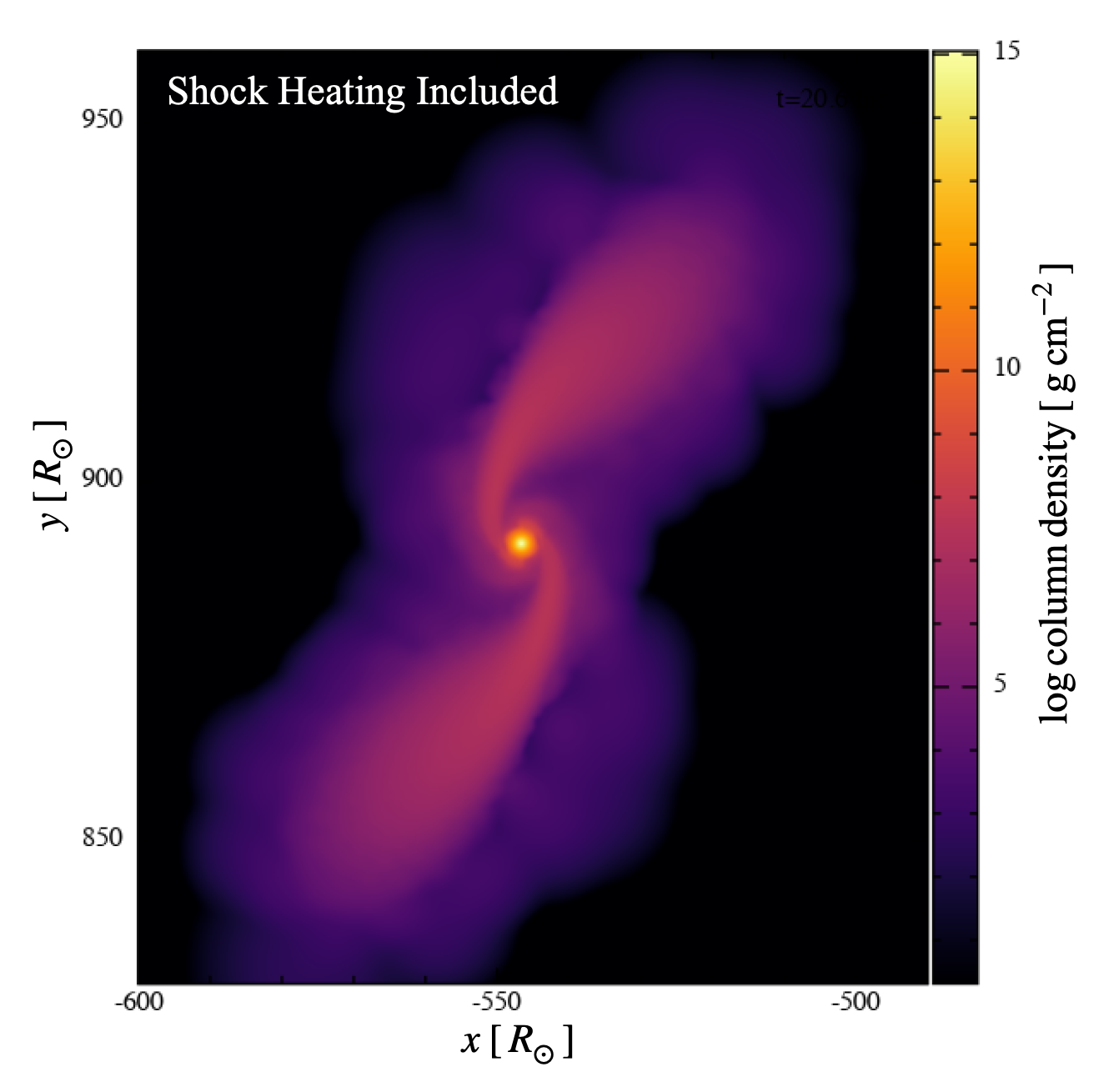}
    \centering
    \includegraphics[height=6.2cm,width=0.43\textwidth]{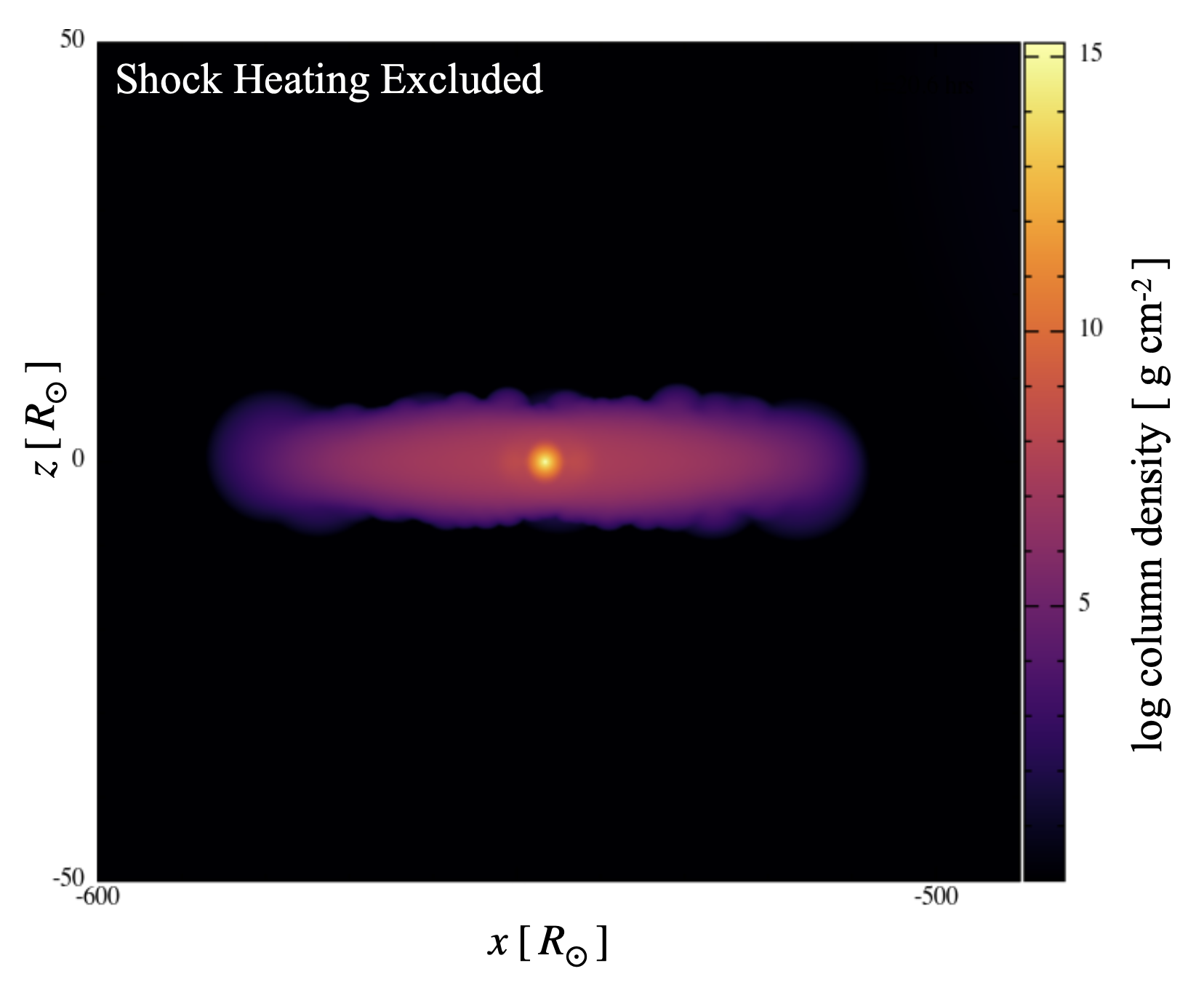} \hspace{5mm}
    \includegraphics[height=6.2cm,width=0.43\textwidth]{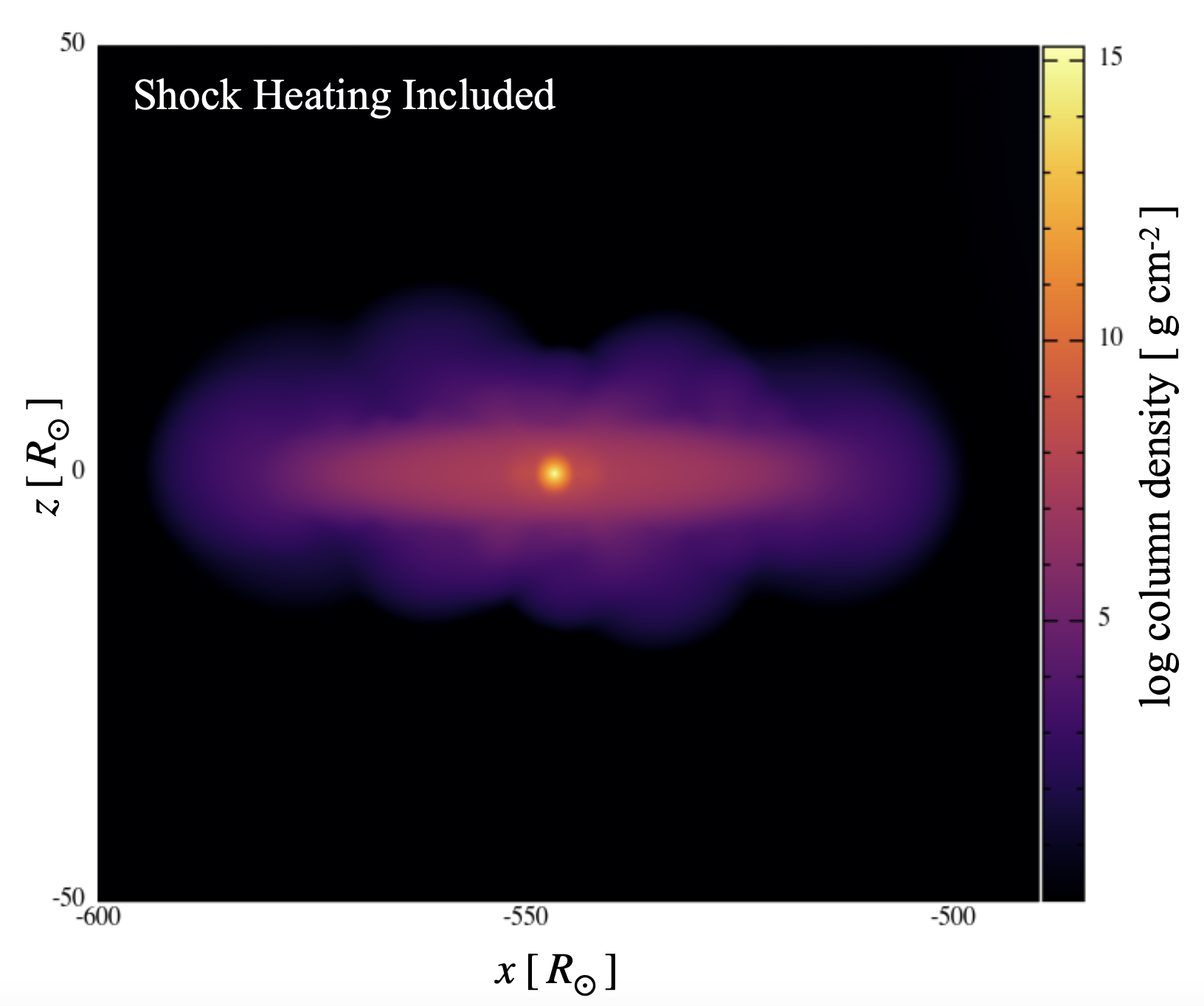}
    
     \caption{The column density profile of the $3 M_{\odot}$ TAMS star at $\sim 1$ day past pericenter passage, with (right) and without (left) shock heating, at a resolution of 10M particles. The top panel shows a comparison of the density profile in $x-y$ coordinates, and the bottom panel shows the density profile in $x-z$ coordinates. The outer layers of the star are inflated in the simulation including shock heating, but structure of the core is largely unaffected, thus preventing the star from being completely disrupted in a few orbits.}
     \label{fig:3msun-density}
\end{figure}

To test the effect of resolution on our results, we also simulated the first pericenter passage of the star with the inclusion of shock heating at 10M particles. Figure~\ref{fig:3msun-shock-heating2} shows the fallback rates with and without shock heating, at the two different resolutions. Aside from differences at the noise level, the curves are identical. We also compared the fallback rates for the $1M_{\odot}$ ZAMS star with $\beta=0.6$ and $\beta=1$, and found no significant differences between the shock heating included and shock heating excluded simulations during the first encounter. Finally, Figure~\ref{fig:3msun-density} shows the fluid column density projected onto ($x$-$y$) and perpendicular to ($x$-$z$) the orbital plane. While the highest-density regions  remain unaltered, the retention of heat causes the outer layers of the star to inflate, which are removed on subsequent encounters. 

Taken together, these tests and the overall insensitivity of our results to the adopted thermodynamic prescription suggest that the star loses the excess heat from tidal dissipation mechanically: it is within the outer layers of the star that most of the thermal energy is deposited, and these layers are subsequently stripped by the black hole. The mechanical (as opposed to radiative) loss of energy is likely required for systems in which the orbital period is as short as ASASSN-14ko's, because the rate at which energy would need to be radiated -- and still yield the observed $\dot{P}$ -- would likely be above the Eddington limit of the star. Specifically, from the relationship between the orbital period $T$ and specific energy $\epsilon$ of a Keplerian orbit, we have $\Delta T/T = 3\Delta\epsilon/(2\epsilon)$, where $\Delta T$ ($\Delta \epsilon$) is the per-orbit change in the orbital period (specific energy). For $T = 114$ days and $M_{\bullet} = 10^7 M_{\odot}$, $\epsilon \simeq 5\times 10^{-4}c^2$, and with $\Delta T/T = 0.001$, we have $\Delta \epsilon \simeq 8.6\times 10^{-7}c^2$. If this energy is placed into the outer $0.01 M_{\odot}$ of the star, then the rate at which the energy would need to be lost over the orbital period of the star is $\dot{E} = 0.01 M_{\odot}\times \Delta\epsilon/T \simeq 1.5\times 10^{39}$ erg s$^{-1}$ $\simeq 3.5 L_{\rm Edd}$, where $L_{\rm Edd} = 4\pi G M_{\star} c/\kappa$ is the Eddington luminosity with $M_{\star} = 3 M_{\odot}$ and $\kappa = \kappa_{\rm es} = 0.34$ cm$^2$ g$^{-1}$ for solar abundances. Since the thermal diffusion timescale is $\sim R_{\star}/(c/\tau) \gg 114$ days, where $\tau$ is the optical depth over the stellar radius, the tidal energy can be deposited and stored in the outer layers of the star (which is also where the amplitude of the tidal force is largest and hence where we expect most of the kinetic energy to be nonlinearly and/or viscously damped), thus allowing the core of the star to survive and giving rise to ASASSN-14o-like flares over a prolonged period.

\begin{table*}[h]
\begin{center}
\begin{tabular}{|c|c|c|}
\hline
Pericenter Passages &Mass Stripped $\left[M_\odot\right]$ &Mass Accreted $\left[M_\odot\right]$ \\ \hline
  & {\bf without shock heating \hspace{2mm} with shock heating} & {\bf without shock heating \hspace{2mm} with shock heating} \\ \hline
 1 & $1.65 \times 10^{-2}$ \hspace{25mm} $1.74 \times 10^{-2}$ & $8.46 \times 10^{-3}$ \hspace{25mm} $8.53 \times 10^{-3}$ \\
 2 & $1.47 \times 10^{-2}$ \hspace{25mm} $2.06\times 10^{-2}$ & $7.54 \times 10^{-3}$ \hspace{25mm} $1.06 \times 10^{-2}$ \\
 3 & $1.31 \times 10^{-2}$ \hspace{25mm} $1.89 \times 10^{-2}$ & $6.45 \times 10^{-3}$ \hspace{25mm} $9.44 \times 10^{-3}$ \\
 4 & $1.19 \times 10^{-2}$ \hspace{25mm} $1.43 \times 10^{-2}$ & $5.95 \times 10^{-3}$ \hspace{25mm} $8.46 \times 10^{-3}$ \\
 5 & $1.12 \times 10^{-2}$ \hspace{25mm} $1.23 \times 10^{-2}$ & $5.55 \times 10^{-3}$ \hspace{25mm} $6.27 \times 10^{-3}$ \\
 6 & $1.07 \times 10^{-2}$ \hspace{25mm} $1.12 \times 10^{-2}$ & $5.37 \times 10^{-3}$ \hspace{25mm} $5.71 \times 10^{-3}$ \\
 \hline
\end{tabular}
\caption{\label{tab:table1} Amount of mass lost per encounter by the $3 M_\odot$ TAMS star orbiting a $10^6 M_\odot$ on a $\beta=1$ orbit, and mass accreted onto the SMBH, with and without the inclusion of shock heating.}
\end{center}
\end{table*}

Finally, we note that the degree to which tidal dissipation modifies the orbital period of the star is a nontrivial and non-monotonic function of the energy. In particular, while tides dissipate orbital energy and further bind the star to the black hole for sufficiently small $\beta$, higher-order moments in the black hole's gravitational field can impart a positive energy to the surviving core once the star loses a sufficiently large amount of mass, as shown by \citet{manukian13, gafton15, cufari23} (see \citealt{kremer22} for an analysis of this phenomenon for stellar-mass black hole encounters). However, these investigations were specific to the case where the star was originally on a parabolic orbit, was not rotating, and had a simple (i.e., polytropic) density profile, none of which apply to the tidal encounters considered here. While we find -- by following the same procedure as outlined in \citet{cufari23} -- that the $3M_{\odot}$ TAMS star on a $\beta = 1$ orbit experiences a reduction in its orbital energy at the level of $0.01 GM_{\star}/R_{\star}$, such that its orbital period would be reduced by an amount that is in rough agreement with the $\dot{P}$ exhibited by ASASSN-14ko (see the preceding paragraph, but we emphasize that our orbit was parabolic, and it is not clear how an initially bound orbit would modify this result), we defer a detailed investigation of this aspect of this problem to future work.

\section{Development of Solid-Body Rotation in the Core}
\label{sec:rotation}
\begin{figure} 
    \centering
    \includegraphics[width=0.9\textwidth]{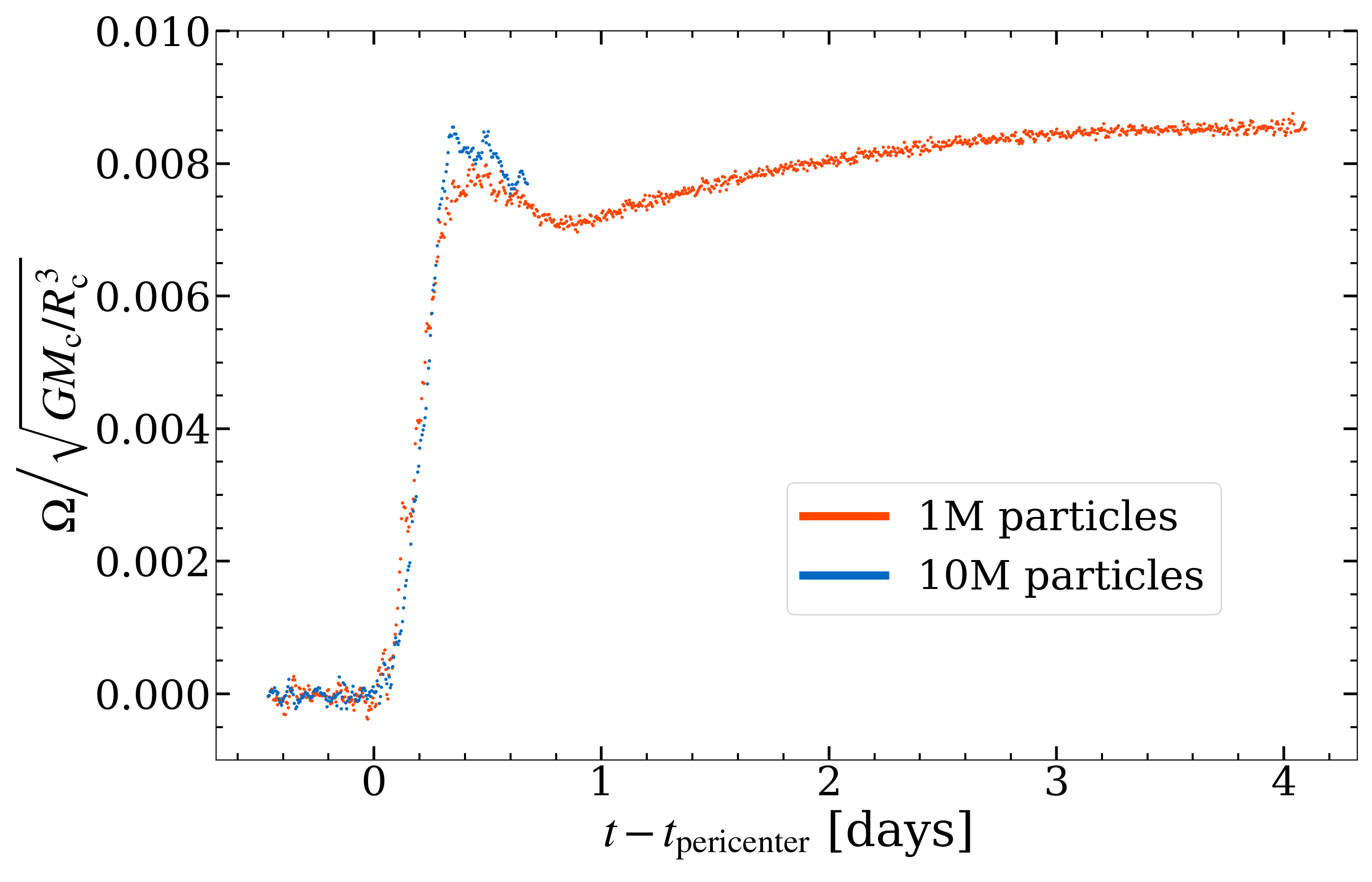}
    \caption{Rotational velocity $\Omega$ imparted to the core of the $3 M_\odot$ TAMS star, in units of its breakup spin, as a function of time measured relative to when the COM of the star reaches pericenter. The initially non-rotating star is spun-up near pericenter, and continues to rotate with a roughly constant angular velocity beyond $\sim 2$ days past pericenter.}
    \label{fig:core-rotation}
\end{figure}

To address the temporal evolution and resolution dependence of the angular velocity of the partially disrupted star, Figure~\ref{fig:core-rotation} shows the angular velocity of the stellar core as a function of time. Here $t_{\rm pericenter}$ is the time taken for a Keplerian orbit to reach pericenter from the initial location of the stellar COM (equal to $5 r_{\rm t}$). The angular velocity $\Omega$ is calculated as an average over every particle present within a radius $R_{\rm c} \sim 0.2 R_{\odot}$ from the center of the star, i.e., within the ``core'' as defined by \citet{coughlin22}. As seen from the figure, the star is spun up to approximately its asymptotic value on $\sim$ the stellar dynamical time upon reaching pericenter, though there is some temporal evolution on longer timescales that likely arises from the re-accretion (by the core) of a fraction of the tidally stripped tails. This figure also shows the angular velocity from the 10M particle simulation to $\sim 1$ day past pericenter, which is in good agreement with the lower-resolution simulation. The solid-body rotation that is induced in the star is therefore resolved and physical.

The emergence of bulk rotation as a consequence of the excitation of quadrupolar modes was studied by ~\cite{kochanek92}, who pointed out that the quadrupolar $\ell=2,m=-2$ mode is tidally excited with a higher amplitude compared to other oscillatory modes, and the conservation of circulation within the star requires that the damping of this mode (which possesses both angular momentum and vorticity) generates a bulk rotation within the star at an angular frequency that is much smaller than the mode frequency (see the discussion in Section 3 of ~\cite{kochanek92}). Our simulations show that tidal interactions spin up the star to a uniform angular velocity, which is a small fraction of $\Omega_{\rm p}$. The rotational energy thus imparted to the core does not exceed its binding energy, and also does not account for the dissipation of orbital energy (as we show in Appendix~\ref{sec:tidal heating} above, the energy imparted through tides heats up the outer layers of the star, which are mechanically removed on subsequent encounters with the SMBH).

\bibliographystyle{aasjournal}

\end{document}